\documentclass[12pt,a4paper]{article}
\usepackage{epsfig,amsmath,amssymb}

\newcommand{\beq}{\begin{equation}}
\newcommand{\eeq}{\end{equation}}
\newcommand{\beqs}{\begin{equation*}}
\newcommand{\eeqs}{\end{equation*}}
\newcommand{\beqa}{\begin{eqnarray}}
\newcommand{\beas}{\begin{eqnarray*}}
\newcommand{\beau}[1]{\begin{equation} \label{#1} \begin{array}{rcl}}
\newcommand{\eeqa}{\end{eqnarray}}
\newcommand{\eeas}{\end{eqnarray*}}
\newcommand{\eeau}{\end{array} \end{equation}}
\newcommand{\bay}{\begin{array}}
\newcommand{\eay}{\end{array}}
\newcommand{\non}{\nonumber}

\newcommand{\ds}{\displaystyle}

\newcommand{\half}{{\frac 1 2}}

\newcommand{\ppt}{{\bf p_t}}
\newcommand{\qq}{{\bf q}}
\newcommand{\rr}{{\bf r}}
\newcommand{\kk}{{\bf k}}
\renewcommand{\P}{{\cal P}}

\newcommand{\Z}{{\cal Z}}

\newcommand{\ra}{{\rightarrow}}

\newcommand{\vev}[1]{\langle #1 \rangle}

\newcommand{\In}[1]{ {{\Big |} _{#1}} }
\newcommand{\Inp}[1]{{{}_{\displaystyle \big| \scriptstyle #1 }} }

\newcommand{\dfp}[2]{\frac{\delta #1}{\delta #2}}

\newcommand{\di}{{\mbox{d}}}

\begin{document}


\begin{titlepage}

\setlength{\textheight}{23.5cm}

\begin{center}
{\Large \bf Minijet transverse spectrum in high-energy hadron-nucleus collisions} \vspace{1cm} \\
        { \bf Alberto Accardi}\footnote{E-mail:{\it accardi@ts.infn.it}}
    and
    { \bf Daniele Treleani}\footnote{E-mail:{\it daniel@ts.infn.it}}
\vspace*{.5cm}  \\
        {\it Dipartimento di Fisica Teorica, Universit\`a di Trieste, \\
        Strada Costiera 11, I-34014 Trieste}  \\
    and \\
    {\it INFN, Sezione di Trieste  \\
        via Valerio 2, I-34127 Trieste }
\vspace{.4cm} \\
\end{center}

\vspace{1cm}
\begin{abstract}
Hadron-nucleus collisions at LHC energies are studied by including
explicitly semi-hard parton rescatterings in the collision
dynamics. Under rather general conditions, we obtain explicit formulae
for the semi-hard cross-section and the inclusive minijet transverse
spectrum.
As an effect of the rescatterings the spectrum is lowered
at small $p_t$ and is enhanced at relatively large transverse
momenta, the deformation being more pronounced at increasing
rapidity.
Its study allows to test the proposed interaction mechanisms and
represents an important baseline to examine nucleus-nucleus collisions.
\end{abstract}

\begin{flushbottom}
\begin{footnotesize}
\centerline{PACS: 11.80.La, 24.85.+p, 25.75.-q}
\end{footnotesize}
\end{flushbottom}

\end{titlepage}

\setcounter{page}{2}
\setcounter{footnote}{0}


\section{Introduction}
\setcounter{equation}{0}

Given the rapid growth of the hard cross-section in hadronic and
nuclear collisions \cite{growth}, the typical inelastic event will
be dominated by the perturbative regime at very high energies so
that, at the LHC, one may expect to be able to derive global
features of the inelastic interaction by perturbative methods.
Such a capability, unavoidably limited to a restricted number of
physical observables, implies however a few non trivial
improvements in the understanding of the mechanisms operating in
the interaction process. To be estimated in a sensible way, different
physical quantities may in fact need a different degree of
understanding of the interaction dynamics, since many details of the
process may be of little relevance for some observables, while 
they may be essential for other quantities. Identifying and
evaluating such physical observables represents a non
trivial improvement in our capability of using the perturbative
QCD to describe physical processes.

An obvious problem will appear when trying to elaborate along
these lines. A perturbative calculation does not introduce any
scale in the dynamics, so that in this case the
kinematic variables are the quantities which give the
dimensionality to the related physical observables. The property
is associated to the relatively low rate of events falling in the
domain of the canonical fixed-$x$ perturbative QCD. On the other
hand the dimensional factor which characterizes the global
features of the typical inelastic event is, rather, the
hadron (or nuclear) scale. When the perturbative regime dominates
a physical observable which represents global features of the
inelastic interaction, the hadron (or nuclear) scale should
therefore appear also in the corresponding perturbative
calculation, presumably introduced through the non-perturbative
input. The structure functions, namely the up to now
non-perturbative input of basically all perturbative calculations,
are on the other hand dimensionless quantities. This implies that
the structure functions, in their present form, will no longer be
an adequate non-perturbative input when trying to accomplish the
program outlined above.

A related aspect is the complexity of the interacting states. The
canonical, fixed-$x$, perturbative QCD approach considers only
perturbative processes initiated by a pair of partons. The
approach is appropriate in the case of very dilute interacting
systems, while it becomes obviously inadequate in a regime with
very large parton densities. In the case of a partonic interaction
in the black disk limit, the initial configuration is in fact
isotropic in transverse space, differently from the final state
produced by an interaction initiated by two partons (namely, at the
leading order in $\alpha_S$, two jets back-to-back in $p_t$), where a
direction in the transverse plane is singled out. A
natural way to recover the black disk symmetry in the final state, is
to include in the interaction perturbative processes
initiated by more than two partons (namely, semi-hard parton
rescatterings), whose relevant property is to
produce many large-$p_t$ jets also at the lowest order in
$\alpha_S$. \\
A non-trivial feature is the associated
non-perturbative input. To deal with processes initiated by more
than two partons one needs in fact to introduce, as a
non-perturbative input, the many-body structure functions, which
contain independent informations on the hadron (or nuclear)
structure with respect to the one-body structure functions needed
to deal with processes initiated by two partons. A basic property is
that the $n$-body structure functions are dimensional quantities,
in such a way that when $n$ is larger than one the many-parton
initiated processes introduce non-perturbative scale factors in
the dynamics of the interaction in a natural way, allowing one to
deal with the problem of dimensionality previously mentioned. \\
By introducing interactions initiated by may partons one may
therefore gain the capability of describing, by means of
perturbative QCD, at least a few general properties of the typical
interaction at the energy of the LHC. To pursue such a program one
should then $i$) evaluate in perturbative QCD processes involving
many partons in the initial state, $ii$) face the problem of the
unknown non perturbative input and develop a strategy in that
respect, and $iii$) study the infrared problem by finding
observable quantities which are infrared stable. This last step
represents the final achievement of the whole program.

The purpose of the present paper is to discuss the case of
hadron-nucleus interactions (hA, for brevity). Being intermediate
between hadron-hadron (hh) and nucleus-nucleus (AA),
hadron-nucleus interactions allow several simplifications in the
formalism developed to discuss heavy-ion collisions. In fact, as
it will be shown hereafter and differently with respect to the
latter case, in the hadron-nucleus instance we will able to obtain
closed analytic expressions for the semi-hard cross-section under
rather general conditions. We will then study the inclusive
minijet transverse spectrum, which is related in a direct way to
the underlying dynamics and is therefore an important baseline for
the study of nucleus-nucleus
collisions.  \\
Beside its intrinsic interest, inclusion of semi-hard rescatterings in
the computation of the transverse spectrum has been advocated
by many authors \cite{TheorCronin,Kastella,Wang1,WW01,Wang2} as the basic
mechanism underlying the Cronin effect \cite{ExpCronin}, namely the
deformation of the hadron $p_t$-spectra in nuclear collisions
as compared with the expectations of a single large-$p_t$ production
mechanism. Multiple parton collisions have also been related to
higher-twist parton distributions
\cite{hightwist0,hightwist,hightwist2}. A non-perturbative study of 
the transverse spectrum in hA collisions in the framework of the
McLerran-Venugopalan model for nuclear and hadronic collisions was
presented in \cite{CGC}.  \\
Another reason of interest in hadron-nucleus collisions is that
theoretical models can be tested against experimental data in a
situation where further nuclear effects are absent, like, e.g., the
formation of a hot and dense medium which can further modify the
transverse spectrum via energy-loss \cite{enloss,Canc}. Therefore a
detailed understanding of hA collisions represents an important
baseline for the generalization to AA collisions
\cite{others,randwalk} and for the discovery of novel physical effects
\cite{QM2001}. \\  

An explicit approach to semi-hard interactions in heavy ion
collisions at the LHC on the lines previously described, has been
accomplished, at least partially, with the help of a few
simplifying hypotheses. The program has been implemented in
\cite{CT90,CT91,CT91b,CT94}, and various physical quantities have
been evaluated in \cite{AT01,Acc01}. The approach relies, to a large
extent,  on the idea of self-shadowing, which we recall for
completeness in Sec.~\ref{ref:selfshad}.
In Sec.~\ref{ref:shxsec} we discuss our expression of the
semi-hard interaction probability between two colliding partonic
configurations. We discuss also the multi-parton distributions,
that are studied with a functional formalism and finally combined
with the interaction probability to derive the hadron-nucleus semi-hard
cross-section. Sec.~\ref{sec:tr.spec.} is devoted to the
discussion of the inclusive minijet transverse spectrum, with
particular emphasis on the mechanism of subtraction of infrared
divergences, which is explicitly implemented in our approach. Results
of numerical evaluations of the inclusive spectra of minijets in
hadron-nucleus collisions are presented in Sec.~\ref{sec:numdisc}. The
last section is devoted to the concluding summary.


\section{Self-shadowing}
\setcounter{equation}{0}
\label{ref:selfshad}

To face the problem of unitarity corrections we make use of the
self-shadowing property of the hard component of the interaction.
For the sake of completeness, in the present paragraph we recall the
main points about the self-shadowing cross-sections in
hadron-nucleus interactions \cite{selfshad}.

Let's consider the inelastic hadron-nucleus cross-section
$(\sigma_{in})_A$, whose expression may be expanded, in
the Glauber approach, as a binomial probability distribution of
inelastic nucleon-nucleon collisions:
\begin{align}
    (\sigma_{in})_A \ &= \int d^2\beta
        \Bigl[1-\Bigl(1-\sigma_{in}\tau(\beta)\Bigr)^A\Bigr]
        \non \\
    &= \int d^2\beta \sum_{n=1}^A \binom{A}{n}
        \bigl(\sigma_{in}\tau(\beta)\bigr)^n
        \bigl(1-\sigma_{in}\tau(\beta)\bigr)^{A-n}
  \label{sinela}
\end{align}
In Eq.~(\ref{sinela}) $\tau(\beta)$ is the nuclear thickness function,
which depends on the impact parameter $\beta$ and is normalized to one, $A$
is the atomic mass number and $\sigma_{in}$ is the inelastic
hadron-nucleon cross-section. One may classify all events
according to a given selection criterion, which we
call ${\cal C}$, while we call ${\cal N}$ the events that are not of
kind ${\cal C}$. We assume that in a hadron-nucleon collision all
events of kind  ${\cal C}$ contribute to $\sigma_{\cal C}$, all other
events contribute to $\sigma_{\cal N}$, so that the inelastic
hadron-nucleon cross-section may be written as
\begin{align*}
    \sigma_{in}=\sigma_{\cal C}+\sigma_{\cal N} \ .
\end{align*}
One may then ask for the expression of the cross-section
$(\sigma_{\cal C})_A$ to produce events of kind ${\cal C}$ in a
collision of a hadron against a nuclear target.
Then, to obtain $(\sigma_{\cal C})_A$, one may
express $(\sigma_{in})^n$ in Eq.~(\ref{sinela}) as a binomial sum
of ``elementary'' events of kind ${\cal C}$ and of kind ${\cal N}$:
\begin{align}
    \sigma_{in}^n = \bigl(\sigma_{\cal C}+\sigma_{\cal N}\bigr)^n
        = \sum_{k=0}^n\binom{n}{k}\sigma_{\cal C}^k
        \sigma_{\cal N}^{n-k} \ .
  \label{sinn}
\end{align}
An interesting case to consider is when the events
of kind ${\cal C}$ are such that any superposition of elementary events of
kind ${\cal C}$, both with events of kind ${\cal C}$ and of kind
${\cal N}$, always gives an event of kind ${\cal C}$. In this case,
all the terms of the sum in Eq.~\eqref{sinn}, with the only exception of
the term with $k=0$, contribute to $(\sigma_{\cal C})_A$, which is
therefore given by: 
\begin{align*}
    (\sigma_{\cal C})_A = \int d^2\beta
        \sum_{n=1}^{A} \binom{A}{n}
        \biggl[\sum_{k=1}^n \binom{n}{k}
        \sigma_{\cal C}^k\sigma_{\cal N}^{n-k} \biggr]
        \bigl(\tau(\beta)\bigr)^n
        \bigl(1-\sigma_{in}\tau(\beta)\bigr)^{A-n} \ .
\end{align*}
By using the relation
\begin{align*}
    \sum_{k=1}^n \binom{n}{k} \sigma_{\cal C}^k \sigma_{\cal N}^{n-k}
        = \sigma_{in}^n-\sigma_{\cal N}^n \ ,
\end{align*}
one obtains:
\begin{align}
    (\sigma_{\cal C})_A \ &= \int d^2\beta \sum_{n=1}^A
        \binom{A}{n} \Bigl[\Big(\sigma_{in}\tau(\beta)\Big)^n
        - \Big(\sigma_{\cal N}\tau(\beta)\Big)^n \Bigr]
        \Bigl[1-\sigma_{in}\tau(\beta)\Bigr]^{A-n}
        \non  \\
    &= \int d^2\beta \Bigl[
        \Bigl(\sigma_{in}\tau(\beta)+1-\sigma_{in}\tau(\beta) \Bigr)^A
        -\Bigl(\sigma_{\cal N}\tau(\beta)+1-\sigma_{in}\tau(\beta)\Bigr)^A
        \Bigr]  \non  \\
    &= \int d^2\beta\Bigl[1-\Bigl(1-\sigma_{\cal C}\tau(\beta)\Bigr)^A\Bigr]
        \non  \\
    &= \int d^2\beta \sum_{n=1}^A \binom{A}{n}
        \Bigl[\sigma_{\cal C}\tau(\beta)\Bigr]^n
        \Bigl[1-\sigma_{\cal C}\tau(\beta)\Bigr]^{A-n} \ .
  \label{selfs}
\end{align}
Notice that, in spite of the fact that we included
superpositions of elementary events of kind $\cal C$ with events
both of kind $\cal C$ and of kind $\cal N$, the nuclear
cross-section $(\sigma_{\cal C})_A$ is obtained by summing all
possible multiple hadron-nucleon interactions of kind ${\cal C}$
alone with a binomial probability distribution, precisely as
$(\sigma_{in})_A$ is obtained by a binomial distribution of
hadron-nucleon inelastic interactions. This relation states the
self shadowing property of the events of kind ${\cal C}$: all
unitarity corrections, namely the term $\bigl[1-\sigma_{\cal
C}\tau(\beta)\bigr]^A$ in the third line of Eq.~\eqref{selfs}, are
expressed by means of the cross-section $\sigma_{\cal C}$ only.
However, this does not mean that $(\sigma_{\cal C})_A$ doesn't
contain events of kind ${\cal N}$, but rather that they are
irrelevant to obtain $(\sigma_{\cal C})_A$. The property that an
event of kind $\cal C$ remains of kind $\cal C$ even after any
number of events of kind $\cal N$ translates into the
disappearance of $\sigma_{\cal N}$ in the nuclear cross-section
$(\sigma_{\cal C})_A$.

Given the discussion above, the only part of the nuclear interaction
that still misses is the cross-section for elementary events of kind
$\cal N$ alone. It can be obtained by considering the following difference
\begin{align}
    \frac{d(\sigma_{in})_A}{d^2\beta}
        -\frac{d(\sigma_{\cal C})_A}{d^2\beta}
        \ &=  \Bigl[1-\sigma_{\cal C}\tau(\beta)\Bigr]^A
        -\Bigl[1-(\sigma_{\cal C }
        +\sigma_{\cal N})\tau(\beta)\Bigr]^A  \non \\
    &= \Bigl[1-\sigma_{\cal C}\tau(\beta)\Bigr]^A
        \times\Biggl\{1-\Biggl[1-\frac{\sigma_{\cal N}\tau(\beta)}
        {1-\sigma_{\cal C}\tau(\beta)}\Biggr]^A\Biggr\}
        \label{bound} \\
    &= \Bigl[1-\sigma_{\cal C}\tau(\beta)\Bigr]^A
        \times \sum_{k=1}^A \binom{A}{k}
        \Biggl(\frac{\sigma_{\cal N }\tau(\beta)}
        {1-\sigma_{\cal C }\tau(\beta)}\Biggr)^k
        \Biggl(1-\frac{\sigma_{\cal N}\tau(\beta)}
        {1-\sigma_{\cal C }\tau(\beta)}\Biggr)^{A-k} , \non
\end{align}
which is therefore bounded by $\bigl[1-\sigma_{\cal
C}\tau(\beta)\bigr]^A$ (second line of \ref{bound}), namely by the
probability of not having any interaction of kind ${\cal C}$ at a
given impact parameter $\beta$. The ratio $\sigma_{\cal N
}\tau(\beta)/[1-\sigma_{\cal C }\tau(\beta)]$ is in fact a quantity
smaller than one, since $\sigma_{in}\tau(\beta)$, which is equal to
$(\sigma_{\cal C}+\sigma_{\cal N})\tau(\beta)$, is a probability. It may be
understood as the probability of an hadron-nucleon interaction
at a given impact parameter,
under the condition that no event of kind ${\cal C }$ takes place.
Hence the last line of Eq.~(\ref{bound}) shows that after removing
all events of kind ${\cal C}$ the interaction is expressed by a
binomial distribution of events of kind ${\cal N}$.

Finally, we observe that if we compute the average number of
hadron-nucleon collisions of kind $\cal C$, $\langle
n\rangle(\sigma_{\cal C})_A$, rather than the cross-section
$(\sigma_{\cal C})_A$, the result is:
\begin{align*}
    \langle n \rangle(\sigma_{\cal C})_A \ &= \int d^2\beta
        \sum_{n=1}^A n \binom{A}{n}
        \bigl(\sigma_{\cal C}\tau(\beta)\bigr)^n
        \bigl(1-\sigma_{\cal C}\tau(\beta)\bigr)^{A-n}
        \non  \\
    &= \int d^2\beta \frac{d}{d\gamma} \sum_{n=1}^A \binom{A}{n}
        \bigl(\sigma_{\cal C}\tau(\beta)\gamma\bigr)^n
        \bigl(1-\sigma_{\cal C}\tau(\beta)\bigr)^{A-n}
        \biggm|_{\gamma=1} \\
    &= A \sigma_{\cal C}
\end{align*}
Notice that the average number of interactions of kind
${\cal C}$ is expressed by the single-scattering term, without any
unitarity correction.


\section{Semi-hard cross-section}
\setcounter{equation}{0}
\label{ref:shxsec}

In this section we want to represent the semi-hard hadron-nucleus
cross-section analogously to the self-shadowing cross-section
\eqref{selfs}, but considering as elementary objects the partons
instead of the nucleons. Indeed, the hard component of the
interaction satisfies the requirements of the self-shadowing
cross-sections if one assumes that a parton which has undergone
interactions with large momentum exchange can always be recognized
in the final state. An immediate difference with respect to the
previous case is that now there is no upper bound on the number of
partons that can take part in the collision. Because of
self-shadowing all unitarity corrections to the semi-hard
cross-section will be therefore expressed by means of the semi-hard
partonic cross-section only, so that one doesn't need to make any
commitment on the soft component when only the semi-hard part of
the interaction is of interest. Self-shadowing allows moreover to
control also the soft component of the interaction by perturbative
means, since that contribution is limited to a fraction of the
cross-section proportional to the probability of not having any
hard interaction at all (see Eq.~\ref{bound}). Obviously the
unavoidable restriction of all considerations done by perturbative
means is that those are limited to partonic final states, whose
properties will hopefully survive hadronization.

To represent the interaction between hadrons and nuclei in terms
of partonic interactions, each one with relatively large momentum
exchange, one needs to write the cross-section for a given
non-perturbative input, namely for a definite partonic
configuration of the two interacting objects. Then, as a perturbative
input, one needs to write the probability of having at least one
semi-hard interactions between the two configurations of partons.
We discuss these two inputs in the next two subsections and in
Sec. \ref{sec:xsec} we combine them to obtain the
hadron-nucleus cross-section.

\subsection{Perturbative input: semi-hard rescatterings}
\label{sec:pertinp}

In a processes involving many
partons in the initial state, an important distinction is
between connected and disconnected hard interactions. A hard
process may in fact be represented also by a disconnected hard
amplitude, since the overall interaction process may still be
connected by the soft component of the interaction. The simplest
case of a disconnected process is obviously the one where the hard
part is represented by two partonic interactions at the lowest
order in perturbative QCD, corresponding to two $2\to2$ parton
scatterings. Since all hard collisions are characterized by short
transverse distances, disconnected hard processes give rise to a
picture of the interaction where the different partonic collisions
are all localized in different points in the transverse space, with a
transverse distance of the order of the scale of soft
interactions \cite{KLL}. The disconnected component of the hard-interaction
leads therefore to a geometrical picture of the process, giving as
well some indications on the degrees of freedom characterizing the
non perturbative input. The many-body parton distributions need in
fact to depend explicitly on the parton transverse coordinates, to
identify the partons involved in each given sub-processes with a
definite localization in transverse space. Notice that the
dependence on the transverse coordinates and the number of partons
taking part the interaction are the basic information needed to
assign the dimensionality to the many-body structure functions,
and therefore to introduce the non-perturbative scale factors in
the interaction dynamics.

While the main feature of the disconnected component of the hard
amplitude is to give rise to a geometrical picture of the
semi-hard interaction, the connected component of the amplitude
becomes more and more structured when one approaches the black
disk limit, where a single projectile parton may interact with
several target partons with large momentum exchange in different
directions in transverse space. The simplest possibility of such
an interaction was discussed in Ref.\cite{CT94}, where the forward
amplitude of the process and all the cuts were derived in the case
of a point-like projectile against two point-like targets, in the
limit of an infinite number of colors and for $t/s\to 0$. In this
case one finds that the different cuts of the $3\to3$ forward
amplitude are all proportional one to another and the
proportionality factors are the AGK weights \cite{agk}. A
consequence is that one may express the three-body interaction as
a product of two-body interaction probabilities. The results
obtained in that simple case may indicate a convenient
approximation of the many-parton interaction probability. One can
in fact argue that the many-parton interaction process may be
approximated by a product of two-parton interactions, so that one
can call the process {\it re-interaction} or {\it rescattering}.
The whole interaction is therefore expressed in terms of two-body
interaction probabilities, precisely as the interaction between
two nuclei is expressed in terms of nucleon-nucleon collisions.
Hence, given a configuration with $n$ partons of the projectile
and $l$ partons of the target, we introduce the probability,
${\cal P}_{n,l}$, of having at least one partonic collision, in a
way analogous to the expression of the inelastic nucleus-nucleus
cross-section \cite{bbc}: 
\begin{align}
    {\cal P}_{n,l}=\Bigl[1-\prod_{i=1}^n
        \prod_{j=1}^l(1-\hat{\sigma}_{ij})\Bigr] \ ,
  \label{prob}
\end{align}
where $\hat{\sigma}_{ij}$ is the probability of
interaction of a given pair of partons $i$ and $j$. Since the
distance over which the hard interactions are localized is much
smaller than the soft interaction scale, one may approximate
$\hat{\sigma}(x_ix_j;b_i-b_j) \approx \sigma(x_ix_j)
\delta^{(2)}(b_i-b_j)$, where $x_i$ and $x_j$ are the momentum
fractions of the colliding partons, $b_i$ and $b_j$ their transverse
coordinates and $\sigma(x_ix_j)$ is the partonic cross
section, whose infrared divergence is cured by introducing a regulator
$p_0$. For example, $p_0$ may be
the lower cutoff on the momentum exchange in each partonic collision,
or a small mass introduced in the transverse propagator to prevent the
divergence of the cross-section at zero momentum exchange. The
expression of ${\cal P}_{n,l}$ is the analogue of Eq.~(\ref{selfs})
and represents the explicit implementation of self-shadowing for the
interaction of two partonic configurations.

\subsection{Non-perturbative input: multi-parton distributions}

In this section we discuss the non perturbative input to the
process. To approach the problem in the most general form we use the
functional formalism introduced in \cite{CT91b}. At a given
resolution, provided by the regulator $p_0$, one may find the nuclear
(or hadronic) system in various
partonic configurations. We call $P^{(n)}(u_1\dots u_n)$ the
probability of a configuration with $n$-partons (the {\it exclusive
$n$-parton distribution}) where $u_i\equiv(b_i,x_i)$ represents
the transverse coordinate of th $i$-th parton, $b_i$, and its longitudinal
fractional momentum, $x_i$. The distributions are symmetric in
the variables $u_i$, and can be obtained from a generating functional
defined with the help of auxiliary functions $J(u)$ as follows:
\begin{align*}
    {\cal Z}[J] = \sum_n{\frac{1}{n!}} \int J(u_1) \dots
        J(u_n) P^{(n)}(u_1,\dots,u_n) du_1\dots du_n \ ,
\end{align*}
all infrared divergences are regularized by $p_0$, which is
implicit in all equations. Probability conservation yields the
normalization condition ${\cal Z}[1] = 1$. Then, the exclusive
$n$-parton distributions can be obtained by differentiating the
generating functional $\cal Z$ with respect to the auxiliary
functions:
\begin{align*}
     P^{(n)}(u_1,\dots,u_n) = \dfp{}{J(u_1)}\dots\dfp{}{J(u_n)}
        {\cal Z}[J]\Inp{J=0} \ .
\end{align*}
A useful representation of $\Z$ may be found by introducing its
logarithm, $\cal F$, with normalization ${\cal F}[1]=0$, so that
\begin{align*}
    {\cal Z}[J]= \, e^{{\cal F}[J]} \ ,
\end{align*}
and by studying the {\it inclusive $n$-parton distribution}, $D^{(n)}$.
They can be obtained as functional derivatives of $\cal Z$ or of $\cal F$.
Indeed
\begin{align*}
    D^{(1)}(u) \ &\equiv P_1(u) + \int P^{(2)}(u,u')du'
        + \half \int P^{(3)}(u,u',u'')du'du''
        + \dots \non \\
    \ &= \frac{\delta{\cal Z}}{\delta J(u)}\biggm|_{J=1}
        = \frac{\delta{\cal F}} {\delta J(u)}{\bigg |}_{J=1}
        \ , \non \\
    D^{(2)}(u_1,u_2) &\equiv P^{(2)}(u_1,u_2)
        + \int P^{(3)}(u_1,u_2,u')du'
        + \half \int P^{(4)}(u_1,u_2,u',u'')du'du''
        \dots \non \\
    &= \frac{\delta^2{\cal Z}}{ \delta J(u_1)\delta J(u_2)}
        {\bigg |}_{J=1} = \frac{\delta^2{\cal F}}
        { \delta J(u_1)\delta J(u_2)}{\bigg |}_{J=1}
        + \frac{\delta{\cal F}}{\delta J(u_1)}
        \frac{\delta{\cal F}}{\delta J(u_2)}{\bigg |}_{J=1} \ ,
        \non
\end{align*}
and so on for higher multi-parton distributions.
These relations show that the correlated part, $C^{(n)}$, of the
inclusive $n$-parton distribution (also called {\it $n$-parton
correlation}) is simply given by differentiation of the generating
functional $\cal F$:  
\begin{align*}
     C^{(n)}(u_1,\dots,u_n) = \dfp{}{J(u_1)}\dots\dfp{}{J(u_n)}
        {\cal F}[J]\Inp{J=1} \ ,
\end{align*}
so that the expansion of ${\cal F}$ near $J=1$ reads:
\begin{align*}
    {\cal F}[J] = \int & \Gamma(u) [J(u)-1] du  \\
        & + \sum_{n=2}^{\infty} \frac{1}{n!}
        \int C^{(n)}(u_1\dots u_n) \bigl[J(u_1)-1\bigr]
        \dots \bigl[J(u_n)-1\bigr] du_1 \dots du_n \ ,
\end{align*}
where $\Gamma(u) \equiv D^{(1)}(u)$ for consistency with the notation
used in previous papers. In this way we have obtained a convenient
representation of the generating functional $\Z = \exp[\cal F]$ in
terms of the single parton inclusive distribution, $\Gamma$, and of the
multi-parton correlations, $C^{(n)}$.
In the simplest case where we neglect all the
correlations between the partons, namely $C^{(n\geq 2)}=0$,
the generating functional is given by
\begin{align}
    {\cal Z}[J] = \, e^{\int \Gamma(u)[J(u)-1]du} \ .
  \label{Znocor}
\end{align}

\subsection{Hadron-nucleus cross-section}
\label{sec:xsec}

The general expression of the semi-hard cross-section at fixed impact
parameter is obtained by
folding the interaction probability, Eq.~(\ref{prob}), with
the multi-parton exclusive distributions of the two colliding systems
(in our case a hadron, $h$, and a nucleus of atomic number $A$):
\begin{align}
    \frac{d\sigma_H}{d^2\beta} =&\ \int \sum_{n=1}^{\infty}
        \frac{1}{n!}
        \dfp{}{J(u_1-\beta)} \dots
        \dfp{}{J(u_n-\beta)} {\cal Z}_h[J]
        \In{J=0} \non \\
    & \times \sum_{m=1}^{\infty} \frac{1}{m!}
        \dfp{}{J'(u_1')} \dots
        \dfp{}{J'(u_m')} {\cal Z}_A[J']
        \In{J'=0} \non \\
    & \times \Bigl\{1-\prod_{i=1}^n\prod_{J=1}^m
        \bigl[1-\hat{\sigma}_{ij}(u,u')\bigr]
        \Bigr\} \prod_{i=1}^{n} du_i \prod_{j=1}^{m} du_j' \ ,
  \label{csc}
\end{align}
where $\beta$ is the impact parameter between $h$ and $A$. To simplify
the notation we introduce the following operators:
\begin{align*}
    \delta_i = \int du_i \dfp{}{J(u_i-\beta)} \ \ ; \ \
    \delta_j' = \int du_i \dfp{}{J'(u_j')} \ .
\end{align*}
Given two functions $f=f(u)$ and $g=g(u)$, the
following identity holds:
\begin{align}
    \, e^{\delta \cdot f} \Z [J+g] = \Z [J+g+f] \ ,
  \label{ident}
\end{align}
where $\delta \cdot f = \int du \delta/\delta J(u) f(u)$.
In other words, the exponential of the operator
$\delta$ acts on the generating functional $\Z$ by shifting its
argument of the amount $f$.

In the case of hadron-nucleus
interactions one may be allowed to neglect the rescatterings of
the partons of the nucleus. Indeed, even at very high center of mass
energies the average number of scattering per incoming parton is
smaller than the average number of nucleons along the parton
trajectory, except in the very forward rapidity region \cite{AT01}.
With this assumption the interaction probability may be simplified as
follows:
\begin{align}
    \Bigl\{1-\prod_{i,j}^{n,m}\bigl[1-\hat{\sigma}_{ij}\bigr]\Bigr\}
        \simeq \sum_{i,j}\hat\sigma_{ij}
        - \frac{1}{2!} \sum_{i,k}\sum_{j\not=l}
          \hat\sigma_{ij}\hat\sigma_{kl}
        + \frac{1}{3!} \sum_{i,k,r} \sum_{j\not=l\not=s}
        \hat\sigma_{ij}\hat\sigma_{kl} \hat\sigma_{rs}+
        \dots
  \label{probnm}
\end{align}
After contracting Eq.~(\ref{probnm}) with the differentiation operators in
Eq.~(\ref{csc}) one obtains
\begin{align*}
    \sum_n \frac{1}{n!}\delta_1\dots\delta_n
        \sum_{q\geq 1} \frac{(-1)^{q-1}}{q!}
        \Bigl[\sum_{i=1}^n \delta'\cdot\hat{\sigma}_i\Bigr]^q
        \, e^{\delta'}
    =& \, \sum_n \frac{1}{n!}\delta_1\dots\delta_n
        \biggl[1-\exp\Bigl(-\sum_{i=1}^n \delta' \cdot
        \hat{\sigma}_i \Bigr) \biggr] \, e^{\delta'} \non \\
    =& \, \biggl\{1-\exp\Bigl[\delta\cdot
        (e^{-\delta'\cdot\hat{\sigma}}-1)
        \Bigr] \biggr\} \, e^{\delta+\delta'} \ . \non
\end{align*}
By using the identity (\ref{ident}), the semi-hard cross-section becomes:
\begin{align}
    \frac{d\sigma_H}{d^2\beta} = \biggl\{1-\exp\Bigl[\delta\cdot
        (e^{-\delta'\cdot\hat{\sigma}}-1)\Bigr]\biggr\}
        \Z_h[J+1] \Z_A[J'+1] \biggm|_{J=J'=0}
  \label{haxsec}
\end{align}
This result is very general and includes all possible parton
correlations of both the projectile and the target; the only
assumption made is that target partons do not suffer any semi-hard
rescattering (we will comment more on this assumption in
Sec.~\ref{sec:modif}). A meaningful approximation (see
Ref.\cite{CT91b}) is to consider the nuclear partons uncorrelated,
namely $C^{(n\geq 2)}_A=0$. Then, by using Eq.~\eqref{Znocor} the
cross-section reduces to:
\begin{align}
    \frac{d\sigma_H}{d^2\beta} \ &=
        \sum_{n=1}^\infty \frac{\delta^n}{n!} \sum_{m=0}^n
        (-1)^{n-m} \binom{m}{n}
        \Z_A[1-m\hat\sigma] \Z_h[J+1]\Inp{J=0} = \non \\
    &= 1-{\cal Z}_h
        \Bigl[e^{-\int\hat{\sigma}(\cdot,u')\Gamma_A(u')du'}
        \Bigr] \ .
  \label{crstot}
\end{align}
If we neglect also the correlations between the partons of the
projectile, we get a further simplification:
\begin{align}
    \frac{d\sigma_H}{d^2\beta} =
        1-\exp\biggl\{-\int du \Gamma_h(u-\beta)
        \Bigl[1-e^{-\int\hat{\sigma}(u,u')\Gamma_A(u')du'}
        \Bigr] \biggr\} \ .
  \label{crstot1}
\end{align}
Both in Eq.~\eqref{crstot} and in Eq.~\eqref{crstot1} the cross-section is a function of
\begin{align}
    W_h(u,\beta) \ &= \Gamma_h(u-\beta)
        \Bigl[1-e^{-\int\hat{\sigma}(u,u')
        \Gamma_A(u')du'}\Bigr] \non  \\
    &= \Gamma_h(u-\beta) {\cal P}_{A}(u) \ ,
  \label{W}
\end{align}
which represents the number of projectile partons that have
interacted with the target, i.e., the projectile {\it wounded
partons} \cite{CT90,CT91b}; we call them {\it minijets}, even if they
did not yet hadronize.
${\cal P}_{A}(u)$ represents the probability that a projectile parton
with given $u=(x,b)$ has at least one semi-hard interaction
with the target, hence the cross-section is obtained by summing all
events with at least one interaction.

One might obtain the average number of wounded partons, Eq.~(\ref{W})
by working out directly from Eq.~(\ref{csc}) the 
average number of projectile partons which have undergone hard
interactions \cite{CT90,CT91b} (a detailed numerical study of this
quantity in Pb-Pb collisions at LHC and RHIC energies is presented
in \cite{AT01,Acc01}). The result, Eq.~(\ref{W}), is obtained
under the only assumption that all the target partons are
uncorrelated. Therefore, $\int du W(u,\beta) =
\vev{n}d\sigma/d^2\beta$ represents the integrated inclusive
cross-section to detect all scattered projectile partons, and
takes into account the correlations of the projectile partons at
all orders. Of course, the projectile parton correlations appear
explicitly in the total hadron-nucleus cross-section. In the
simplest case of two-parton correlations one would obtain:
\begin{align}
    \frac{d\sigma_H}{d^2\beta} = 1 - & \exp\biggl\{
        - \int du  W_h(u-\beta)
         + \frac1{2!} \int dudu' {\cal P}_{A}(u)
        C_h^{(2)}(u-\beta,u'-\beta) {\cal P}_{A}(u')
        \biggl\} \ .
  \label{xseccor}
\end{align}
The effect of correlations on $d\sigma_H/d^2\beta$ is however small,
both when unitarity corrections are small 
(i.e., when the semi-hard parton-parton cross-section is small, so
that  $\P_A$ and $W_h$ are both of order $\sigma_H$) and when they are
large (i.e., when $\sigma_H$ is large, ${\cal P}_{A} \sim 1$ and $W_h$ 
is large). If, on the other hand, one is looking for correlations,
the simplest quantity which depends linearly on $C_h^{(2)}$ is the
double-jet inclusive cross-section.

\section{Inclusive minijet transverse spectrum}
\setcounter{equation}{0}
\label{sec:tr.spec.}

The whole semi-hard hadron-nucleus cross-section results from the
superposition of the multiple interactions of the partons of the
projectile hadron (which is a dilute partonic system) with the
nuclear target (which is a dense partonic system). The inclusive
transverse spectrum of the projectile minijets is given by the
distribution in $p_t$ of the average number of wounded partons,
and is affected by the presence of semi-hard rescatterings
\cite{CT91}.
The deformation of the high-$p_t$ hadron spectra, which leads to the
Cronin effect, was studied in terms of semi-hard parton
rescatterings in \cite{TheorCronin,Kastella,Wang1,WW01}, where partons
that suffered up to two scatterings where included, leading to a good
description of the data for pA collisions up to $\sqrt{s} = 39$
GeV/A. However, the two-scattering approximation
breaks down at higher energies, except at very high $p_t$, and the
whole wounded parton transverse spectrum is needed.
More phenomenological approaches
\cite{Wang2,others}, which take into account also
intrinsic transverse momentum, model the effects of multiple
scattering as an additional Gaussian $p_t$-broadening for each
rescattering suffered by a parton. A random-walk model of the multiple
scatterings was proposed in \cite{randwalk}.

After the introduction of semi-hard parton rescatterings, integrated
quantities like the semi-hard cross-section and the  minijet
multiplicity show a weak dependence on the infrared cutoff needed to
regularize the infrared divergences arising in the perturbative
computations \cite{CT90,AT01}. On the contrary, it will be shown that
differential quantities like the minijet $p_t$-spectrum are more
sensitive on the detailed dynamics of the interaction and show a
stronger dependence on the cutoff, if only logarithmic. To reduce
this dependence on the cutoff one needs to improve further the picture
of the dynamics by including also gluon radiation in the interaction
process. Some steps along this line in the case of deep inelastic
electron-nucleus scattering have been presented in \cite{hightwist2}.
In this paper, however, we neglect the problem of the gluon radiation
and we concentrate on the effects of elastic rescatterings.

\subsection{Transverse spectrum}

We can expand the average number of projectile wounded partons,
Eq.~\eqref{W}, at a given $x$ and $b$ in a
collision with impact parameter $\beta$, in the following way:
\begin{align}
    W_h(x,b,\beta) = \Gamma_h(x,b-\beta) \sum_{\nu=1}^{\infty}
        \frac{\langle n_A(x,b)\rangle^\nu}{\nu!}
        \, e^{-\langle n_A(x,b)\rangle} \ ,
  \label{W1}
\end{align}
where $\langle n_A(x,b)\rangle\equiv \int dx'\Gamma_A(x',b)
\sigma(xx')$ is the average number of
scatterings of a projectile parton at a given $x$ and $b$
\cite{CT91b}. The average number of wounded partons is then
given by the average number of incoming partons, $\Gamma_h$,
multiplied by the probability of having at least one semi-hard
scattering, which is given by a Poisson distribution in the
number of scatterings, $\nu$, with average number $\vev{n_A(x,b)}$.
Therefore, we can obtain the inclusive differential distribution in
$p_t$ by introducing a constraint in the transverse
momentum integrals that give the integrated parton-parton cross
sections in the expression above:
\begin{align}
    \frac{dW_h}{d^2p_t}(x,b,\beta)
        \ = &\ \Gamma_h(x,b-\beta) \sum_{\nu=1}^{\infty}
        \frac{1}{\nu!} \int \Gamma_A(x'_1,b) \dots
        \Gamma_A(x'_{\nu},b)
        \, e^{ -\int dx'\Gamma_A(x',b) \sigma(xx')} \,
        \non \\
    & \times \, \frac{d\sigma}{d^2k_1} \dots \frac{d\sigma}{d^2k_{\nu}}
        \, \delta^{(2)}({\bf k}_1 + \dots + {\bf k}_{\nu}
        -{\bf p_t})
        \,\, d^2k_1 \dots d^2k_{\nu} \,\, dx_1' \dots dx_{\nu}' \ .
  \label{dWdp}
\end{align}
The limits of integration on $x'_i$ and $x'$ are
respectively $xx_i's\geq 4k_i^2$ and $xx's\geq 4p_0^2$, and all
the distribution functions are evaluated for simplicity at a fixed
scale.

By using the above formula one can study the $p_t$-broadening of a
wounded parton, namely, the square root of the average transverse
momentum squared acquired through its path across the nucleus.
Consider a single projectile parton with fixed $x$ and $b$. The
probability that it acquires a certain $p_t$ after the collision is
given by Eq.~\eqref{dWdp} divided by the number,
$\Gamma_h(x,b-\beta)$, of incoming partons:
\begin{align*}
    \frac{d{\cal P}_A(x,b)}{d^2p_t} = \frac{dW_h}{d^2p_t}(x,b,\beta)
        \frac{1}{\Gamma_h(x,b-\beta)} \ .
\end{align*}
Then, the average transverse momentum squared of a wounded parton is given by
\begin{align*}
    \vev{p_t^2(x,b)}_A = \frac{\vev{\vev{p_t^2}}}{\vev{\vev{1}}} \ ,
\end{align*}
where $\vev{\vev{f(p_t)}} = \int d^2p_t f(p_t) d{\cal P}_A/d^2p_t$.
The delta function in  Eq.~\eqref{dWdp} tells that at a fixed number,
$\nu$, of scatterings we have
$\vev{\vev{p_t^2}}_\nu=\vev{\vev{(\sum_{i=1}^\nu
\kk_i)^2}}=\vev{\vev{\sum_{i=1}^\nu k_i^2}}$. The last equality is due
to the azimuthal symmetry of the differential parton-parton
cross-sections $\frac{d\sigma}{d^2k_i}$. The resulting expression is
symmetric under exchanges of the transverse momenta $k_i$, so that
$\vev{\vev{p_t^2}}_\nu = \nu\vev{\vev{k_i^2}}$. Then it is
immediate to see that
\begin{align}
    \vev{p_t^2(x,b)}_A = \frac{1}{{\cal P}_A }
        \int d^2p_t dx' p_t^2
        \frac{d\sigma}{d^2p_t}(xx') \Gamma_A(x',b)
        = \vev{p_t^2(x,b)}_1
        \frac{\vev{n_A(x,b)}}{{\cal P}_A(x,b)}  \ ,
  \label{ktbroad}
\end{align}
where
\begin{align*}
    \vev{p_t^2(x,b)}_1 = \frac{\int d^2p_t dx' p_t^2
        \frac{d\sigma(xx')}{d^2p_t} \Gamma_A(x',b)}
        {\int dx' \sigma(xx') \Gamma_A(x',b)}
\end{align*}
is the average transverse momentum squared in a single parton-parton collision.
The $p_t$-broadening of the wounded partons in a hA collision is
then given by the $p_t$-broadening in a single collision multiplied by
the average number of rescatterings suffered by a wounded parton.
A similar result for the $p_t$-broadening of a fast parton traversing
a nuclear medium was derived in \cite{JKT01}. The $p_t$-broadening was
also studied in different contexts in \cite{hightwist,hightwist2}.\\
Two interesting limits can be considered:
\begin{align}
    \vev{p_t^2(x,b)}_A \sim \left\{
        \bay{ll}
          \vev{p_t^2(x,b)}_1 & \mbox{as\ } p_0 \ra \infty  \\
          \vev{p_t^2(x,b)}_1 \vev{n_A(x,b)}
            & \mbox{as\ } p_0 \ra 0 \ .
        \eay \right.
  \label{limits}
\end{align}
Since the minijet yield is dominated by transverse momenta of the
order of the cutoff, these two limits say roughly that the minijets at
high $p_t$ (i.e., high $p_0$ in Eq.~\eqref{limits}) suffer mainly one
scattering. On the contrary, at low $p_t$ (i.e., low $p_0$ in
Eq.~\eqref{limits}) they undergo a random walk
in the transverse momentum plane and the broadening is proportional to
the average number of steps in the random walk,
i.e., the average number of semi-hard scatterings suffered along the
wounded parton trajectory. This picture will be studied in more
detail in Sec.~\ref{sec:modif}.

An explicit formula for the transverse spectrum can be
obtained by studying its Fourier transform, since all the convolutions
in Eq.~\eqref{dWdp} turn into products and the sum over $\nu$ may be
explicitly performed. To this purpose, we introduce the Fourier
transform of the parton-parton scattering cross-section
\begin{align*}
    \tilde{\sigma}(v;xx') = \int d^2k \, e^{i{\bf k}\cdot{\boldsymbol v}}
        \frac{d\sigma}{d^2k}(xx') \ .
\end{align*}
Note that $\tilde{\sigma}(0;xx') = \sigma(xx')$ and that due to the
azimuthal symmetry of $d\sigma/d^2k$, its Fourier transform depends
only on the modulus, $v$, of ${\boldsymbol v}$. Then, the transverse
spectrum \eqref{dWdp} may be written as:
\begin{align}
    \frac{dW_h}{d^2p_t}(x,b,\beta) =
        \Gamma_h(x,b-\beta)
        \int \frac{d^2v}{(2\pi)^2}
        \, e^{-i{\ppt\cdot{\boldsymbol v}} \widetilde W_h(v;x,b) } \ ,
  \label{trsp}
\end{align}
where
\begin{align}
    \widetilde W_h(v;x,b) \ & = \  \sum_{\nu=1}^{\infty} \frac{1}{\nu!}
        \left[\int dx' \Gamma_A(x',b)
        \tilde{\sigma}(v;xx') \right]^{\nu}
        \, e^{-\int dx' \Gamma_A(x',b)
        \tilde{\sigma}(0;xx')} \non \\
    & =  e^{\int dx' \Gamma_A(x',b)
        \{\tilde{\sigma}(v;xx')-\tilde{\sigma}(0;xx')\}}
        - e^{-\int dx' \Gamma_A(x',b)
        \tilde{\sigma}(0;xx')} \, .
  \label{fttrsp}
\end{align}
An immediate consequence is that the transverse spectrum has a finite
limit as $p_t \ra 0$, even when a cutoff on the momentum exchange is
used:
\begin{align*}
    \frac{dW_h}{d^2p_t}\In{\ppt=0} & (x,b,\beta)
        =  \Gamma_h(x,b-\beta) \int \frac{d^2v}{(2\pi)^2}
          \widetilde W_h(v;x,b)  \ .
\end{align*}

\subsection{Expansion in the number of scatterings}
\label{sec:expscat}

We can obtain an expansion of $\widetilde W_h$ in the number of the
rescatterings suffered by the incoming parton by expanding
Eq.~\eqref{fttrsp} in powers of $\tilde\sigma$:
\begin{align}
    \widetilde W_h & (v;x,b) \ =
        \sum_{\nu=1}^\infty \widetilde W^{(\nu)}_h(v;x,b)
        \non \\
    & =  \sum_{\nu=1}^\infty \frac{1}{\nu !}
        \Big[ \Big( \int dx' \Gamma_A(x',b)
        \big[\tilde{\sigma}(v;xx')-\tilde{\sigma}(0;xx')\big]
        \Big)^\nu - \Big( - \int dx' \Gamma_A(x',b)
         \tilde{\sigma}(0;xx') \Big)^\nu \Big] \ .
  \label{Wexp}
\end{align}
Coming back to the $p_t$ space, the expansion of the transverse
spectrum in number of scatterings reads:
\begin{align}
    \frac{dW_h}{d^2p_t}(x,b,\beta)
        =  \sum_{\nu=1}^\infty
        \frac{dW_h^{(\nu)}}{d^2p_t}(x,b,\beta)
        = \sum_{\nu=1}^\infty
        \Gamma_h(x,b-\beta) \int \frac{d^2v}{(2\pi)^2}
        \, e^{-i{\bf p}_t\cdot{\boldsymbol v}} \widetilde
        W^{(\nu)}_h(v;x,b) \ .
  \label{ftexp}
\end{align}
The series Eq.~\eqref{Wexp} can be obtained also by expanding
$\widetilde W(v)$ around $v=0$. Since the variable $v$ is
Fourier-conjugated to $p_t$, the expansion of the transverse spectrum,
Eq.~\eqref{ftexp}, will be valid at high $p_t$ and we expect a
breakdown of any truncation at sufficiently low momentum.
Note that we can obtain this high-$p_t$ expansion of the spectrum
directly in $p_t$ space by expanding the exponential in (\ref{dWdp})
and collecting the terms of the same order in $\sigma$.
As an example, the first three terms, Eqs. \eqref{1sc}, \eqref{2sc} and
\eqref{3sc}, can be found in the appendix. The first two, suitably
symmetrized, were used in \cite{TheorCronin,Kastella,Wang1,WW01} to
explain the Cronin effect up to $\sqrt s = 39$ GeV/A.
The study of this series is the subject of Sec.~\ref{sec:cancdiv};
numerical results up to $n=3$ scatterings will be discussed in
Sec.~\ref{sec:modif} and compared to the whole spectrum.
In the appendix we will discuss the symmetrization of the terms
of the series.

\subsection{Cancellation of the divergences}
\label{sec:cancdiv}

All terms of the expansion (\ref{ftexp}) are divergent in
the infrared region so that we need to cure them with the
regulator $p_0$.
Nevertheless, the infrared divergences are already regularized to a
large extent by the subtraction terms
originated by the expansion of $\exp[-\langle n_A(x,b)\rangle]$
appearing in Eq.~\eqref{dWdp}, namely by the constraint of probability
conservation. This cancellation mechanism was observed also  in
Ref.\cite{Kastella} for the two-scattering term and in Ref.\cite{Canc}
in a different context.

It is instructive to look in detail how the subtraction works for the
lower order terms of the expansion. We start by considering the case
of a single rescattering ($\nu=2$). To simplify the notation we
write the elementary differential cross-section $d\sigma/d^2k$ as
$\sigma(\kk)$, and notice that it depends only on the modulus, $k$, of the
momentum.  By expressing the semi-hard cross-section as $\sigma=\int
d^2k \sigma(\kk)$ the term of order $\sigma^2$ may be written as
\begin{align}
    \frac{dW_h^{(2)}}{d^2p_t}(x,b,\beta) =& \ \Gamma_h(x,b-\beta)
        \int \Gamma_A(x'_1,b) \Gamma_A(x'_2,b) dx'_1dx'_2
        d^2k_1 d^2k_2
  \label{dW2} \\
        &\times \frac{\sigma(\kk_1) \sigma(\kk_2)}{2}
        \Bigl[\delta^{(2)}({\bf k}_1+{\bf k}_2-{\bf p}_t)
        - \delta^{(2)}({\bf k}_1-{\bf p}_t)
        - \delta^{(2)}({\bf k}_2-{\bf p}_t) \Bigr]
        \ , \non
\end{align}
where the first term in the square brackets represent two successive
scatterings with no absorption. The two negative terms are the
corrections induced by the expansion of the absorption factor
$\exp[-\langle n_A(x,b)\rangle]$ of the single-scattering term,
$\nu=1$ in (\ref{dWdp}), and correspond to a single-scattering along
with the effects of absorption in the initial or final state.
The expression we obtained is symmetric in the integration variables
${\bf k}_1$ and ${\bf k}_2$. \\
The cutoff dependence is originated by the singular behavior of
the integrand for ${\bf k}_1\approx0$ or for ${\bf k}_2\approx0$,
since the $\delta$-functions in the square brackets prevent
the possibility of ${\bf k}_1$ and ${\bf k}_2$ being both zero at the
same time. Because of the symmetry under the exchange
${\bf k}_1\leftrightarrow{\bf k}_2$, to study the cutoff
dependence of Eq.~(\ref{dW2}) it is enough to discuss the integration
around ${\bf k}_1 = 0$. In the region ${\bf k}_1\approx0$ the
term $\delta^{(2)}({\bf k}_1-{\bf p}_t)$ does not contribute, as long as
${\bf p}_t$ is finite. The integration in ${\bf k}_2$ is done with
the help of the $\delta$-functions and one obtains
\begin{align*}
    \int d^2k_1 \sigma(\kk_1)
        \Bigl[ \sigma(\ppt-{\bf k}_1)-\sigma(\ppt) \Bigr]
        \ .
\end{align*}
On the other hand, for ${\bf k}_1\approx0$, one may use the expansion
\begin{align*}
    \sigma(\ppt-{\bf k}_1) \simeq \sigma(\ppt)-\sigma'(\ppt)
        \frac{{\bf p}_t\cdot{\bf k}_1}{p_t} \ ,
\end{align*}
where ${\bf p}_t\cdot{\bf k}_1$ represents the scalar
product of the two vectors, and $\sigma'(\ppt)=\frac{d}{d|\ppt|}\sigma(\ppt)$
depends only on the modulus of $\ppt$. One is left with
\begin{align*}
    -\frac{\sigma'(\ppt)}{p_t} \int {\bf p}_t\cdot{\bf k}_1
        \sigma(\kk_1)d^2k_1=0 \ ,
\end{align*}
where the vanishing result is due to the azimuthal
symmetry of $\sigma(\kk_1)$. The dominant contribution to the
integral comes therefore from the next term in the expansion of
$\sigma({\bf k}_1-{\bf p}_t)$, which goes as $k_1^2$. Hence the
resulting singularity is only logarithmic in $p_0$, since
$\sigma(\kk) \sim k^{-4}$ as $k\to0$.
The subtraction terms, originated by the absorption factor $\exp[-\langle
n_A(x,b)\rangle]$ in Eq.~(\ref{dW2}), have cancelled the singularity
of the rescattering term almost completely. This feature is common
to all the terms of the expansion (\ref{dW2}). 

Hereafter we consider in detail the term with two rescatterings:
\begin{align}
    \frac{dW_h^{(3)}}{d^2p_t}(x,b,\beta) = &\ \Gamma_h(x,b-\beta)
        \int \Gamma_A(x'_1,b) \Gamma_A(x'_2,b) \Gamma_A(x'_3,b)
        dx'_1dx'_2dx'_3 d^2k_1 d^2k_2 d^2k_3
        \non \\
        &\times \frac{\sigma(\kk_1)\sigma(\kk_2)\sigma(\kk_3)}{6}
        \Bigl[\delta^{(2)}({\bf k}_1 +{\bf k}_2+{\bf k}_3-{\bf p}_t)
        \non \\
    &- \delta^{(2)}({\bf k}_1+{\bf k}_2-{\bf p}_t)
        -\delta^{(2)}({\bf k}_2+{\bf k}_3-{\bf p}_t)
        -\delta^{(2)}({\bf k}_3+{\bf k}_1-{\bf p}_t)
        \non \\
    &+ \delta^{(2)}({\bf k}_1-{\bf p}_t)
        + \delta^{(2)}({\bf k}_2-{\bf p}_t)
        +\delta^{(2)}({\bf k}_3-{\bf p}_t)\Bigr] \ .
  \label{dW3}
\end{align}
The different $\delta$-functions in Eq.~(\ref{dW3})
correspond to all the terms of order $\sigma^3$ in Eq.~(\ref{dWdp})
and represent the triple scattering term together with all
subtraction terms, induced by the expansion of the absorption
factor $\exp[-\langle n_A(x,b)\rangle]$ of the double- and of the
single-scattering terms. The expression
has been symmetrized with respect to ${\bf k}_1$, ${\bf k}_2$ and
${\bf k}_3$ and is singular for ${\bf k}_1=0$, ${\bf
k}_2=0$ and ${\bf k}_3=0$.  The $\delta$-functions in \eqref{dW3}
prevent the tree momenta to be close to zero at the same time, then
we start by discussing the most singular
configuration corresponding to two integration variables both
close to zero. Given the symmetry of the integrand it is enough to
study the integration region with ${\bf k}_1\approx0$, ${\bf
k}_2\approx0$. In this region the terms $\delta^{(2)}({\bf k}_1+{\bf
k}_2-{\bf p}_t)$, $\delta^{(2)}({\bf k}_1-{\bf p}_t)$ and $\delta^{(2)}({\bf
k}_2-{\bf p}_t)$ do not contribute. The integrals on the
transverse momenta are therefore written as
\begin{align}
    \int d^2k_1 d^2k_2 \sigma(\kk_1) \sigma(\kk_2)
        \Bigl[\sigma({\bf p}_t-{\bf k}_1-{\bf k}_2)
        - \sigma({\bf p}_t-{\bf k}_1)
        -\sigma({\bf p}_t-{\bf k}_2)
        +\sigma(\ppt) \Bigr] \ .
  \label{dblsing}
\end{align}
In the region ${\bf k}_1\approx0$, ${\bf k}_2\approx0$
one may use the expansion
\begin{align}
    \sigma(\ppt-{\bf k}) \simeq & \sigma(\ppt)-\sigma'(\ppt)
        \frac{{\ppt\cdot{\bf k}}}{p_t} + \frac{1}{2}
        \Biggl[\sigma''(\ppt)\frac{(\ppt
        \cdot{\bf k})^2}{p_t^2}-\sigma'(\ppt)\
        \frac{(\ppt\times{\bf k})^2}{p_t^3} \Biggr] \ ,
  \label{expans}
\end{align}
where $\ppt\times{\bf k}$ represents the vector
product of $\ppt$ and ${\bf k}$ and $\sigma''(\ppt) = \frac{d^2}{d|\ppt|^2}
\sigma(\ppt)$ depends only on the modulus of $\ppt$. All terms
proportional to 
$\sigma(\ppt)$ cancel and all the terms linear in ${\bf k}$ integrate
to zero thanks to the azimuthal symmetry of $\sigma(\kk)$. Then
one is left with
\begin{align*}
    \int d^2k_1 d^2k_2 \sigma(\kk_1) \sigma(\kk_2)
        & \Biggl\{ \frac{\sigma''(\ppt)}{2p_t^2}
        \Bigl[\bigl({\bf p}_t\cdot({\bf k}_1
        +{\bf k}_2)\bigr)^2
        -\bigl({\bf p}_t\cdot{\bf k}_1\bigr)^2
        -\bigl({\bf p}_t\cdot{\bf k}_2\bigr)^2\Bigr] \\
    & -\frac{\sigma'(\ppt)}{2p_t^3} \Bigl[\bigl(
        {\bf p}_t\times({\bf k}_1+{\bf k}_2)\bigr)^2
        -\bigl({\bf p}_t\times{\bf k}_1\bigr)^2
        -\bigl({\bf p}_t\times{\bf k}_2\bigr)^2 \Bigr]
        \Biggr\} \ ,
\end{align*}
which simplifies to
\begin{align*}
    \int d^2k_1 d^2k_2 \sigma(\kk_1) \sigma(\kk_2)
        \Biggl\{ \frac{\sigma''(\ppt)}{p_t^2}
        \bigl({\bf p}_t\cdot{\bf k}_1\bigr)
        \bigl({\bf p}_t\cdot{\bf k}_2\bigr)
        -\frac{\sigma'(\ppt)}{p_t^3}
        \bigl({\bf p}_t\times{\bf k}_1\bigr)
        \bigl({\bf p}_t\times{\bf k}_2\bigr) \Biggr{\}} = 0 \ .
\end{align*}
The result is again zero because of the azimuthal
symmetry of $\sigma(\kk)$. Hence, all terms of the expansion
(\ref{expans}) up to the second order in $k$ do not contribute.
All other terms linear in ${\bf k}_1$ or in ${\bf
k}_2$, which are obtained from the first terms in the square brackets in
Eq.~(\ref{dblsing}), do not contribute for the same reason, so the
first term different from zero is at least of order $k_1^2k_2^2$,
and originates a square-logarithm singularity as a function of the
regulator $p_0$. 

One may repeat the argument for the regions where only one of the
integration variables is close to zero. We consider in detail the
case ${\bf k}_1\approx0$ and ${\bf k}_2$, ${\bf k}_3$ both finite.
In this region the term $\delta^{(2)}(\ppt-{\bf k})$ does not
contribute to Eq.~(\ref{dW3}). The transverse momentum integrals are
therefore
\begin{align*}
    \int d^2k_1 d^2k_2 \sigma(\kk_1)\sigma(\kk_2)
        & \Bigl\{ \sigma({\bf p}_t-{\bf k}_1-{\bf k}_2)
        -\sigma({\bf p}_t-{\bf k}_1)
        -\sigma({\bf p}_t-{\bf k}_2)
        +\sigma(\ppt) \Bigr{\}} \\
    & + \int d^2k_1 d^2k_3 \sigma(\kk_1) \sigma(\kk_3)
        \Bigl\{ -\sigma({\bf p}_t-{\bf k}_1)
        +\sigma(\ppt) \Bigr\} \ .
\end{align*}
To study the singularity it is sufficient to keep the
first two terms in the expansion of $\sigma({\bf k})$ in \ref{expans}.
One obtains
\begin{align*}
    \int & d^2k_1 d^2k_2 \sigma(\kk_1) \sigma(\kk_2) \\
    & \times \Biggl\{ \sigma({\bf p}_t-{\bf k}_2)
        -\sigma'({\bf p}_t-{\bf k}_2)
        \frac{({\bf p}_t-{\bf k}_2)\cdot{\bf k}_1}
        {{\bf p}_t-{\bf k}_2} -\sigma(\ppt)
        +\sigma'(\ppt)\frac{{\bf p}_t\cdot{\bf k}_1}{p_t}
        -\sigma({\bf p}_t-{\bf k}_2)
        +\sigma(\ppt)\Biggr{\}}  \\
    & + \int d^2k_1 d^2k_3 \sigma(\kk_1) \sigma(\kk_3)
        \Biggl\{ -\sigma(\ppt)
        + \sigma'(\ppt)\frac{{\bf p}_t\cdot{\bf k}_1}{p_t}
        - \sigma(\ppt) \Biggr{\}} \ ,
\end{align*}
which simplifies to
\begin{align*}
    \int d^2k_1 d^2k_2 \sigma(\kk_1) \sigma(\kk_2)
        \Biggl\{ -\sigma'({\bf p}_t-{\bf k}_2)
        \frac{({\bf p}_t-{\bf k}_2)\cdot{\bf k}_1}
        {{\bf p}_t-{\bf k}_2} + 2\sigma'(\ppt)
        \frac{{\bf p}_t\cdot{\bf k}_1}{p_t} \Biggr\} = 0 \ .
\end{align*}
As in the previous case one obtains a vanishing result thanks to the
azimuthal symmetry of $\sigma(\kk)$.
All integrations in the singular points induce therefore
at most a square-logarithm singularity, as a function of the cutoff, in
the term with $\nu=3$ in Eq.~(\ref{ftexp}).

The argument holds for the whole spectrum, as one may see by
looking at its Fourier transform, Eq.~(\ref{fttrsp}). Indeed,
to study the dependence of the inclusive spectrum on the regulator $p_0$
at a given $p_t$ different from zero one needs
to consider the first term in the square brackets only. The
cutoff enters in the difference
\begin{align*}
    \tilde{\sigma}(v)-\tilde{\sigma}(0)
        = \int \frac{d\sigma}{d^2k}
        \Bigl[e^{i{\bf k}\cdot{\boldsymbol v}}-1\Bigr]{d^2k}
        =-v^2\frac{\pi}{2}\int_{p_0}^{\infty}k^3
        \frac{d\sigma}{d^2k}dk+{\rm \ finite\ terms}
\end{align*}
so that, also in this case, the divergence for $p_0\to0$ is only
logarithmic.


\section{Numerical results and discussion}
\setcounter{equation}{0}
\label{sec:numdisc}

In this section we discuss in detail, both qualitatively and quantitatively,
the modifications induced by the rescatterings on the minijet
inclusive transverse spectrum. We consider a proton-lead
collision with center of mass energy $\sqrt s = 6$ TeV/A and impact
parameter $\beta = 0$. 
In the numerical computations we used the leading order perturbative
parton-parton cross-section with a mass regulator $m\equiv p_0$:
\begin{align*}
    \frac{d\sigma}{d^2p}(xx') = k \frac{9\pi\alpha_s(Q)^2}{(p^2+m^2)^2}
        \theta\big(xx's-4(p^2+m^2)\big)
\end{align*}
where $k$ is the $k$-factor that simulates next-to-leading order
corrections (we chose $k=2$). The single-parton nuclear
distribution function has been taken to be factorized in $x$ and $b$:
\begin{align*}
    \Gamma_A(x,b) = \tau_A(b) G(x,Q)
\end{align*}
where $\tau_A$ is the nuclear thickness function and $G$ is the proton
distribution function.  We evaluated the strong coupling
constant and the nuclear distribution functions at a fixed scale $Q=m$.
In the computations we used a hard-sphere
geometry
\begin{align*}
    \tau_A(b) = A \frac{3}{2\pi R^3} \sqrt{R^2-b^2}\theta(R^2-b^2)
\end{align*}
where $R=1.12 A^{1/3}$ is the nuclear radius. For $G$ we used the
GRV98LO parameterization \cite{GRV98}.
At low $p_t$ the spectrum is obtained by computing numerically the
Fourier transform in Eq.~(\ref{trsp}), but
at high $p_t$ the result begins to oscillate too much, and in that
region the spectrum was computed by using the expansion in
the number of scattering up to the three-scattering term (the formulae
actually used, Eqs. \eqref{1sc}, \eqref{2scsym}
and \eqref{3scsym}, are discussed in the appendix). We checked that
the spectrum obtained by Fourier transformation matched smoothly the
expansion.

\subsection{Effects of rescatterings}
\label{sec:modif}

In this section we discuss the projectile and the target transverse
spectrum averaged over a given rapidity interval:
\begin{align}
    \frac{dW_{h}}{d^2p_t}(\beta,\eta_{min},\eta_{max})
        = \frac{1}{\eta_{max}-\eta_{min}}
        \int_{\eta \in [\eta_{min},\eta_{max}]}
        \hspace*{-1.4cm} dx d^2b \hspace*{.3cm}
        \frac{dW_h}{d^2p_t}(x,b,\beta) \ ,
  \label{projspec}
\end{align}
where we approximated the pseudo-rapidity by $\eta=\log(x\sqrt s/p_0)$.
The target spectrum, $dW_{A}/d^2p_t$, is obtained by
interchanging $h$ and $A$ in Eq.~\eqref{projspec}. Note that now we
are taking into account all possible rescatterings of the target, as well.

\begin{figure}[t]
\begin{center}
\epsfig{figure=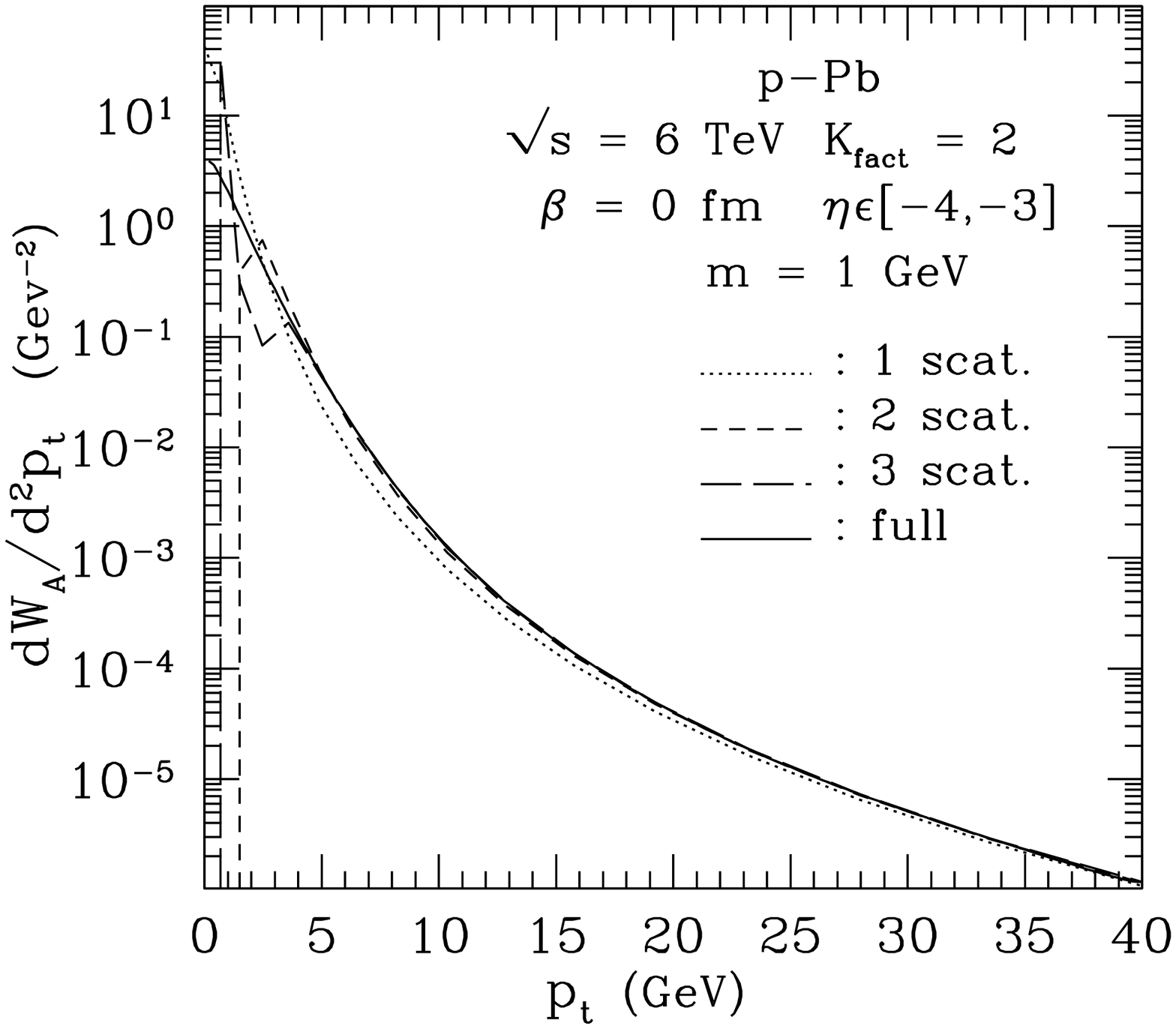,height=7.9cm}
\epsfig{figure=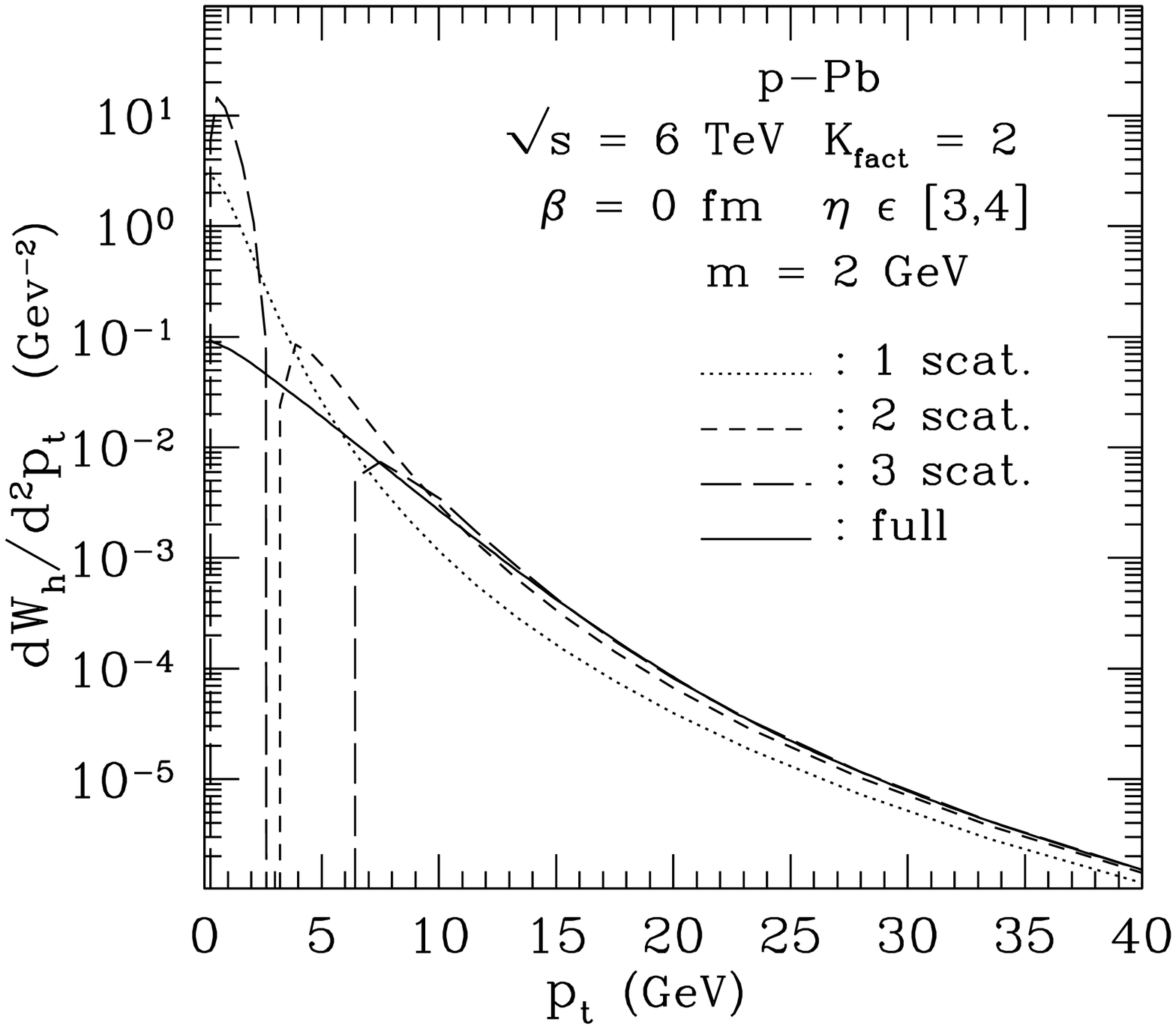,height=7.9cm}
 \caption{\footnotesize
{\it Left:} Target $p_t$-spectrum for $\eta \in [-4,-3]$.
{\it Right:} Projectile $p_t$-spectrum for $\eta \in [3,4]$.
The full transverse spectrum (solid line) is compared with the one-,
two- and three-scattering approximations (viz., dotted, short-dashed
and long-dashed lines).
  \label{fig:expansion}  }
\end{center}
\end{figure}

In fig.\ref{fig:expansion} we compare the full transverse
spectrum (solid line) with its expansion in the number of scatterings
up to three scatterings (dotted and dashed lines). We show both the
projectile and target minijet spectrum in a pseudo-rapidity
region $\eta \in [3,4]$ for the projectile and $\eta \in [-4,-3]$ for
the target. Note that the rapidity is defined with reference to the
projectile hadron direction of motion. The choice of a forward region
(backward for the target) is done to
enhance the effect of the rescatterings and to better discuss the
deformation induced in the spectrum. Indeed, in those regions
the average fractional momentum of an incoming parton is large, so
that the number of available target partons is large and the
probability of rescattering becomes large.

First, we look at the projectile spectrum.
At high $p_t$ the spectrum is enhanced with respect to the single
scattering approximation because of the $p_t$ broadening induced by
the rescatterings. As $p_t$ is further increased it approaches the
single-scattering spectrum, as expected on general grounds when the
$p_t$ distribution of the elementary scattering follows a power law.
This can be understood qualitatively by looking at the path
in $p_t$-space followed by the incoming parton. Given a final large
$p_t$, due to the leading divergences in Eq.~(\ref{dW2}), the
leading processes to get that $p_t$ with two semi-hard scatterings
are a first scattering with momentum transfer $q_1 \approx p_0$
followed by a second one with $q_2 \approx p_t$ and vice-versa. For an
analogous reason, the leading configuration to reach the final
$p_t$ with three scatterings is $q_1 \approx p_t$ plus $q_2 \approx q_3
\approx p_0$ and permutations. This sequence of three
scatterings is less probable than the process with two scatterings as
$p_t$ increases because the fraction of phase-space volume that this
process occupies decreases much faster with $p_t$ than in the
two-scattering case. For an analogous reason also
the relative importance of the two-scattering term with respect to the
single-scattering term decreases as $p_t$ increases.
In conclusion as $p_t$ increases the average number of scatterings per
parton decreases, and eventually the spectrum is well described by the
single-scattering approximation.

At intermediate $p_t$ the average number of scatterings per parton
increases and the shape of the spectrum is more and more distorted
with respect to the single-scattering case. In fact, the fraction of
phase-space available to the leading configuration of a multiple
scattering process ($q_1 \approx p_t$, $q_2 \approx \dots \approx q_n
\approx p_0$ and permutations) increases as $p_t$ decreases.
However, this is not the only mechanism at work. Indeed, in our
computation each wounded parton is
counted as one minijet in the final state, independently of the number
of rescatterings. On the other hand, in the single-scattering
approximation one identifies the number of minijets in the final state
with the number of parton-parton collision. This leads to an
overestimate of the jet multiplicity and to a divergence of the
spectrum at $p_t=0$ as $p_0$ goes to zero. Therefore at low $p_t$ the
minijet yield is more and more suppressed with respect to the single
scattering approximation.

At very low transverse momentum $p_t \lesssim p_0$ a parton
undergoes a large number of rescatterings, all with $q_i\approx
p_0$. Hence, the parton is doing a random-walk in the transverse plane
and the spectrum becomes flat as $p_t \ra 0$ because the phase space
becomes isotropically populated. This shows that at very low $p_t$
multiple semi-hard scatterings are consistent with the random-walk
model of Ref. \cite{randwalk}, while at moderate and high-$p_t$ the
physical picture is rather different.

By comparing the results for the projectile and target transverse
spectrum one sees that a projectile parton is traversing a very dense
target and the effects of the rescatterings are large. On the
contrary, a target parton sees a rather dilute system, and its minijet
spectrum does not differ too much from the single-scattering result,
except at very low $p_t$. Moreover the changes induced by the
rescatterings on integrated quantities, like those entering in the
expression of the hadron-nucleus cross-section, are minimal. This
is consistent with our approximation of not including rescatterings for
the target partons to obtain analytic formulae for the
hadron-nucleus cross-section. One can also see that the
three-scattering approximation describes well the projectile spectrum
for $p_t \gtrsim 15$ GeV, while it breaks down completely at $p_t
\lesssim 7$ GeV, where it becomes negative. For the target spectrum the
three-scattering approximation is not accurate for $p_t \lesssim 4$ GeV.

\subsection{Minijet inclusive transverse spectrum}
\label{sec:mjspec}

In this section we study the minijet transverse spectrum resulting
from the sum of the transverse spectra of the projectile and target
wounded partons:
\begin{align}
    \frac{dW_{hA}}{d^2p_t}(\beta,\eta_{min},\eta_{max})
        = \frac{1}{\eta_{max}-\eta_{min}}
        \int_{\eta \in [\eta_{min},\eta_{max}]}
        \hspace*{-1.4cm} dx d^2b \hspace*{.3cm}
        \Big( \frac{dW_h}{d^2p_t}(x,b,\beta)
        + \frac{dW_A}{d^2p_t}(x,b,\beta) \Big) \ .
  \label{totspec}
\end{align}
We analyze the spectrum in three rapidity regions, namely $\eta \in
[-4,-3]$, $\eta \in [-1,1]$ and $\eta \in [3,4]$ (respectively
``backward'', ``central'' and ``forward'' with reference to the
projectile direction of motion). While the target partons basically do
not suffer any rescattering in all three regions, the projectile partons
undergoes many rescatterings in the forward region, some in the
central region and basically no one backwards.

\begin{figure}[t]
\begin{center}
\vskip -1cm
\hspace*{-1.1cm}
\epsfig{figure=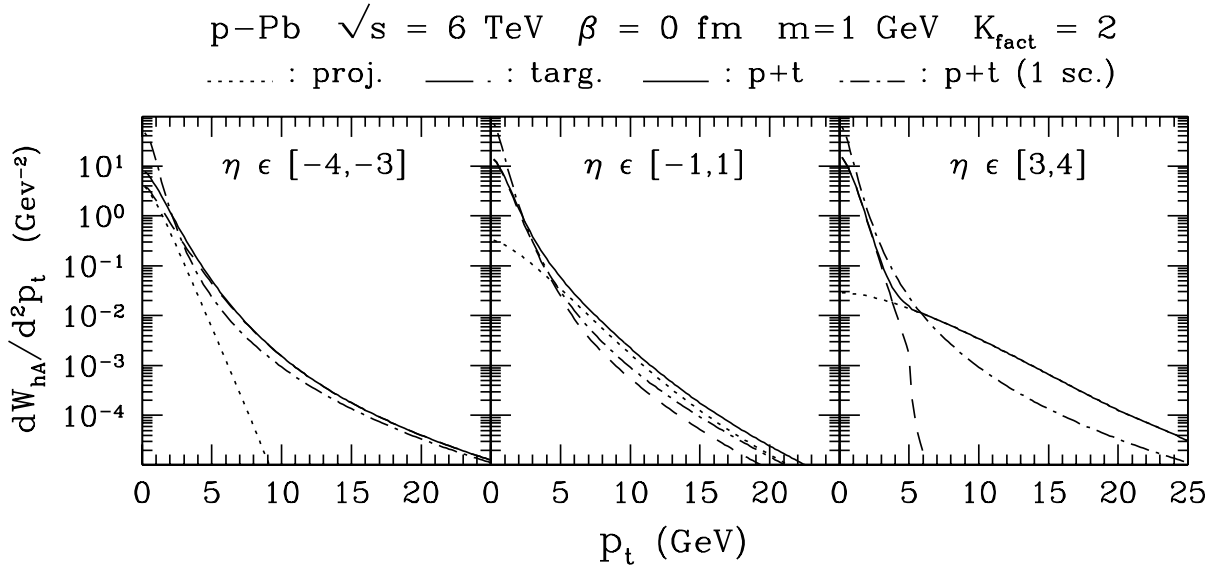,width=19cm}
\vskip -10.5cm
\caption{\footnotesize
Projectile plus target  $p_t$-spectrum (solid line) at different
rapidities compared to the result of the one-scattering approximation
(dot-dashed line). Also shown are the contributions of the projectile
minijets (dotted line) and and of the target minijets (dashed line).
  \label{fig:spectot}  }
\end{center}
\end{figure}

In Fig. \ref{fig:spectot} we show the spectrum \eqref{totspec} (solid
line) and the contributions of the projectile and of the target (dotted
and dashed lines, respectively). For comparison also the total
spectrum obtained in the one-scattering approximation is plotted
(dot-dashed line). The spectra are computed with a regulating mass
$m=1$ GeV. \\
In the backward region both the projectile and the target suffer
mainly one scattering over all the $p_t$-range except at $p_t \sim 0$, and
the spectrum is dominated almost everywhere by target minijets. \\
In central and forward
regions the target jets still suffer basically one scattering over all
the $p_t$ range. On the contrary, the projectile crosses a denser and
denser target and undergoes an average number of rescatterings that
increases with pseudo-rapidity. This means that at low $p_t$ the
projectile spectrum is very reduced with respect to the one-scattering
approximation, and the minijet yield may become negligible with
respect to the minijet yield from target.
The overall effect is that at low $p_t$ the spectrum is
dominated by minijet production from the target while at intermediate
and high $p_t$ it is dominated by minijet production from the
projectile. \\
At very forward rapidities this effect becomes quite dramatic and the
spectrum acquires a structured shape: it follows the inverse power
behaviour of the single-scattering term at high $p_t$, it is concave
at intermediate $p_t$ because of the suppression of the projectile minijets
and becomes convex again at low $p_t$, where the target begins to dominate.

\begin{figure}[t]
\begin{center}
\vskip -1cm
\hspace*{-1.1cm}
\epsfig{figure=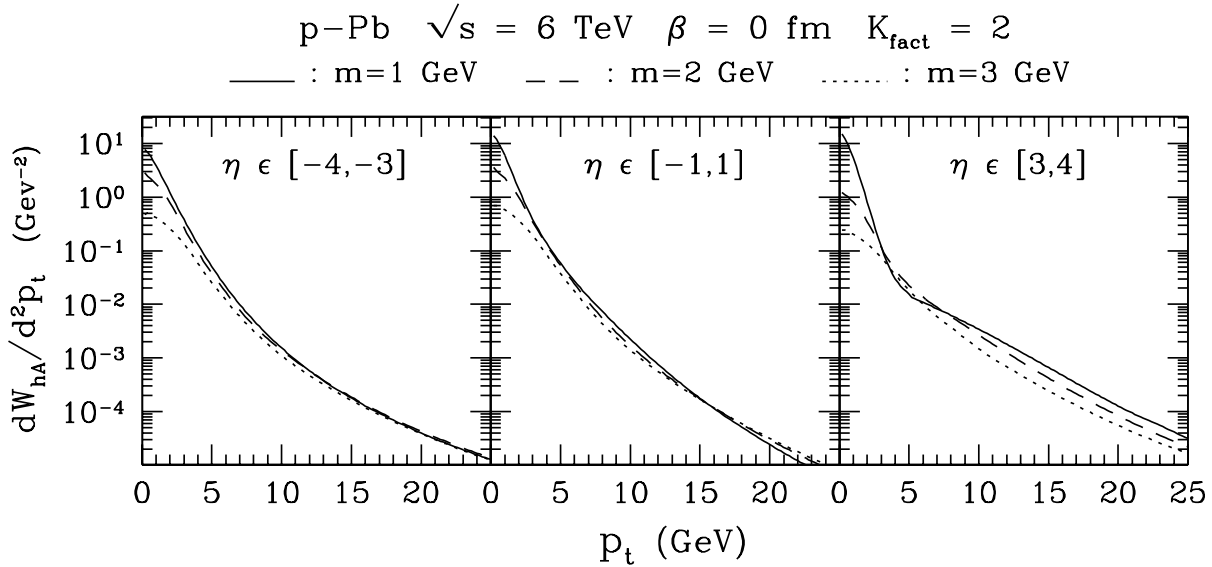,width=19cm}
\vskip -10.5cm
\caption{\footnotesize
Regulator dependence of the projectile plus target $p_t$-spectrum  at
different rapidities for $m=1,2,3$ GeV (viz., solid, dashed and dotted line).
  \label{fig:speccfr}  }
\end{center}
\end{figure}

In Fig.\ref{fig:speccfr} we study the dependence of the
spectrum on the choice of the cutoff, and plotted the result for $m =
1,2,3$ GeV. The deformation of the spectrum decreases as the regulator
increases (indeed, the average number of rescattering decreases) and
for $m \gtrsim 3$ GeV it begins to become negligible.

\begin{figure}[t]
\begin{center}
\vskip -1cm
\hspace*{-1.6cm}
\epsfig{figure=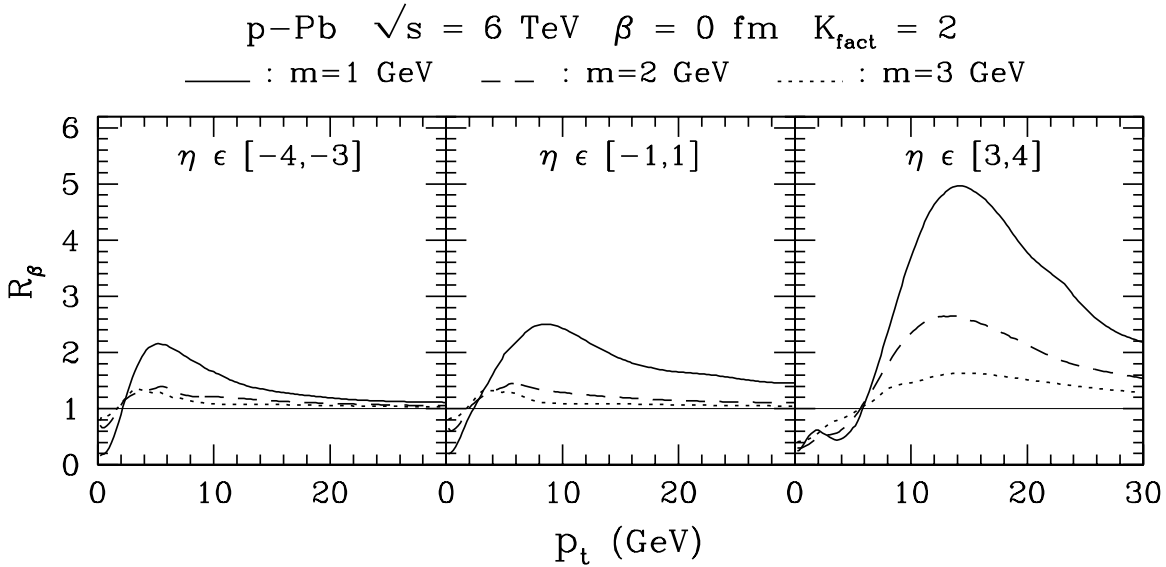,width=19cm}
\vskip -10.5cm
\caption{\footnotesize
Ratio of the full projectile plus parton $p_t$-spectrum to the
one-scattering approximation at different rapidities and for
$m=1,2,3$ GeV (viz., solid, dashed and dotted line).
  \label{fig:ratio}  }
\end{center}
\end{figure}

The effects of the rescatterings are better displayed by studying the
ratio of the full transverse spectrum and the single-scattering
approximation:
\begin{align}
    R_\beta(p_t) = \frac{dW_{hA}/d^2p_t}{dW^{(1)}_{hA}/d^2p_t}
        = \frac{dW_{hA}/d^2p_t}{A_\beta \ dW^{(1)}_{pp}/d^2p_t}
        \ ,
  \label{Rbeta}
\end{align}
where $A_\beta \simeq \int d^2b \tau_h(b-\beta)\tau_A(b)$
is the number of target nucleons interacting with the projectile
at a given impact parameter.  \\
In Fig.\ref{fig:ratio} we plotted the ratio $R_\beta(p_t)$, which
measures the Cronin effect for minijet production, computed with three
different regulators $m=1,2,3$ GeV.
At $m=3$ GeV the effect of the rescatterings is rather small in all
the three rapidity intervals, except at very low $p_t$, and doesn't affect
the integrated quantities like the average number of minijets.
As the regulating mass is decreased the rescatterings begin to show
up, and lead to a big effect in the forward region.

The ratio $R_\beta(p_t)$ is carachterized by three quantities: the
momentum $p_\times$ where the $R_\beta$ crosses 1, the
momentum $p_M$ where it reaches the maximum and the height $R_M$ of
the maximum. \\
The sensitivity of $p_\times$ on the cutoff decreases as the
pseudo-rapidity increases. Loosely speaking, when the average number
of scatterings is high, as it is the case at $p_t \simeq p_\times$,
the jets loose memory of $p_0$, which gives the order of magnitude of
the typical momentum exchanged in each collision. $p_M$ shows a
slightly larger sensitivity on the regulator, since it lies in a
region where the average number of scatterings is smaller.
This behaviour is very
different from the conclusions drawn by considering only the
expansion up to two scatterings, where both $p_\times$ and $p_M$ are
proportional to $p_0$ \cite{WW01}. In fact, at low center of mass
energies the two-scattering is a good approximation in all
rapidity ranges, except may be very forward. However, it breaks down in
any case at transverse momenta comparable to the regulator $p_0$.
Therefore, while most of the spectrum is well described by the
two-scattering approximation, the behaviour of $p_\times$ and $p_M$ is
not.  \\
On the other hand, the height of the peak is much more
sensitive to the cutoff, since its leading term is roughly
proportional to some power of the logarithm of the regulator:
\begin{align*}
    \left[ \frac{dW_{hA}}{d^2p_t} - \frac{dW^{(1)}_{hA}}{d^2p_t}
        \right]_{p_t=p_M} \underset{p_0 \ra 0}{\sim}
        \left[ \log\left(\frac{p_M^2}{p_0^2}\right)
        \right]^{\vev{n_{resc}(p_M)}} \non
\end{align*}
Since $p_M$ is not very large, the average number of rescatterings at that
value of the transverse momentum, $\vev{n_{resc}(p_M)}$, is much greater
than one and the sensitivity of $R_M$ on $p_0$ is high. At high
$p_t$ the average number of rescatterings tends to zero, so the
sensitivity of the $R_\beta$ on $p_0$ decreases and disappears at very
large transverse momenta. \\
Note that the peak is located in a $p_t$-region, where soft
interactions (which have been disregarded in our approach)
are expected to be negligible, therefore in that region our perturbative
computations should describe almost completely the spectrum.
Following Ref. \cite{WW01}
we might interpret $p_0$ as the momentum scale at which
the interaction deviates from the perturbative computations. With this
interpretation $p_0$ would acquire a physical meaning: though
physics doesn't know about the artificial subdivision in hard and soft
interactions, it is a well defined question to ask up to what scale
are the perturbative computations good. If the collision dynamics would be
determined by parton multiple elastic scatterings alone, then the
measure of the height of the peak would be a way of measuring $p_0$. \\
On the other hand, the sensitivity of $R_\beta$ on $p_0$
is rather signaling a lack in our description of the dynamics underlying
the hadron-nucleus collision. We expect that such a
sensitivity will be considerably reduced when including in the dynamics
also the gluon radiation emitted by the multiply scattering partons.
Some of the effects of the radiation on
the transverse spectrum might be however described by the
parameter $p_0$ in the model where the radiation is neglected.
Since the inclusion of gluon
radiation in the dynamics would introduce new physical scales,
like the radiation formation time, related to the energy of the
collision and the nuclear size, we would expect in any case that the
value of $p_0$ will depend on $\sqrt s$ and $A$.


\section{Conclusions}
\setcounter{equation}{0}

The purpose of the present article is to draw the attention to
some of the advantages of studying hadron-nucleus semi-hard
interactions at the LHC. As in the case of lower energies, hA
interactions represent an important intermediate step to relate hh
and AA reactions, being much simpler to understand as compared
with the latter. Moreover, even at higher energies, like those
obtainable at RHIC and LHC, in hA collisions we don't expect the
formation of a dense and hot system, like the quark-gluon plasma,
so that one can study directly the nuclear modification of the
dynamics without the need of disentangling the effects of the
structure of the target and those due to the formation and
evolution of the dense system. Hadron-nucleus interactions
represent therefore the baseline for the detection and the study
of the new phenomena peculiar to AA collisions.

We faced the problem of unitarity corrections to the semi-hard
cross-section by including explicitly semi-hard parton rescatterings in
the collision dynamics, exploiting the self-shadowing property of
the semi-hard interactions. In the interaction mechanism we took
into account just elastic parton-parton collisions, while we
neglected the production processes at the partonic level (e.g.,
all $2\to3$ etc. elementary partonic
processes), whose inclusion represents a non-trivial step in our approach
and deserves further study. \\
Contrary to the case of AA collisions, we have been able to obtain
closed analytic expressions for the semi-hard hA cross-section,
Eq.~(\ref{crstot}). To that purpose a crucial assumption has been
to consider the hadron as a dilute system, so that rescatterings
of nuclear partons can be neglected, while rescatterings of the
projectile are fully taken into account. In our expressions we
have disregarded correlations in the nuclear multi-parton
distributions, whose effect may be nevertheless studied in a
straightforward way within the present functional approach. \\
We have then focused on the inclusive minijet transverse spectrum at
fixed impact parameter, Eq.~(\ref{trsp}), which is influenced in a more
direct way by the rescatterings. The modifications of the transverse
spectrum induced by the semi-hard rescatterings of the projectile
partons is emphasized in the ratio $R_{\beta}(p_t)$,
Eq.\ref{Rbeta}, defined as our $p_t$ spectrum divided by the impulse
approximation. In particular, we have evaluated it at $\beta=0$ for
different values of the regulator $p_0$. The results
are described by the values of $p_\times$ (defined by
$R_{\beta}(p_\times)=1$), $p_M$ (which is the value of $p_t$ that
maximizes the ratio) and $R_M$ (which is the maximum of
$R_{\beta}$). We obtain the both $p_\times$ and $p_M$ depend
weakly on $p_0$, while $R_M$ has, on the contrary, a strong
dependence on $p_0$ also when the regulator is rather
small. Therefore, the results
for the spectrum allow also to identify the limits of
the picture of the dynamics considered in this paper.
Analogously to the 
average transverse energy and the number of minijets in AA
collisions \cite{AT01}, some of the features of $R_\beta$, like
$p_\times$ and $p_M$, show a tendency towards a limiting value
at small $p_0$. All these quantities depend therefore only
marginally on details of the dynamics which have not been taken
into account in the present approach. On the contrary, the limits
of the simplified picture of the interaction show up in $R_M$.
Because of the strong dependence of $R_M$ on $p_0$, in order to
describe the spectrum one needs in fact to fix experimentally the
value of $p_0$ by measuring $R_M$. This feature might be not so
unpleasant, because if one limits the analysis to the inclusive transverse
spectrum of minijets in hA collisions, all the effects which are not
taken into account in the interaction (like the gluon radiation in the
elementary collision process) are summarized by the value of a single
phenomenological parameter. However this feature will not hold any
further if one had to evaluate more differential
properties of the produced state, which can be properly discussed
only after introducing explicitly further details in the description
of the elementary interaction process. \\
The experimental measure of the Cronin effect in minijet
production in hA collisions would be therefore of major importance: it would
allow to establish the correctness of the whole approach here
described and it would represent the basis for a 
deeper insight in the semi hard interaction dynamics both for hA
and AA collisions.

{\bf Acknowledgements.} A.A. would like to thank B.~Kopeliovich,
I.~Lokhtin, A.H.~Muel\-ler, L.~McLerran, A.~Polleri, U.~A.~Wiedemann, and
F.~Yuan for their comments and many useful discussions. This work was
partially supported by the Italian Ministry of University and of 
Scientific and Technological Research ({\sc MURST}) by the grant 
{\sc COFIN99}.


\newpage
\begin{appendix}

\section{Expansion in the number of collisions}
\setcounter{equation}{0}

For the numerical computation of the high-$p_t$ expansion of the
minijet spectrum in the number of scatterings suffered by a projectile
parton it is convenient to implement the subtraction of the IR
divergences directly in the integrand.
In this way the Monte-Carlo integrations, which we use
because of the high dimensionality of the phase space (in particular
for three or more scatterings), work at their best.
In fact, Eqs.~\eqref{dW2} and \eqref{dW3} are not suited for
numerical implementation due to the delta functions. The basic
property that allowed the cancellation of the
divergence in the integrand was the symmetry under exchanges of the
integration variables. Unfortunately after using the delta-functions
to perform on of the integrals, one obtains in general
non-symmetric expressions.

The goal of this appendix is to study how to symmetrize each term of
the expansion of the transverse spectrum. We will discuss them in
detail up to the three-scattering term, but the techniques discussed
can be applied also to the generic term in the expansion.
For simplicity, we will use the following
notation, already introduced in the main text:
\begin{align*}
    \sigma({\bf k}) = \frac{d\sigma}{d^2k}(xx') \ .
\end{align*}

\subsection{One-scattering term}

The one-scattering term doesn't include any subtraction term, so that
we don't need to symmetrize it. It is simply given by
\begin{align}
    \frac{dW_h}{d^2p_t}^{(1)} (x,b,\beta) =  \Gamma_h(x,b-\beta)
        \int  dx' \Gamma_A(x',b) \sigma(\ppt) \ ,
  \label{1sc}
\end{align}
and corresponds to the result one obtains by considering just
disconnected parton collisions and neglecting parton rescatterings. It
corresponds also to modeling the hadron-nucleus collision as a
superposition of hadron-nucleus collisions.

\subsection{Two-scattering term}

The two-scattering term is given by Eq.~(\ref{dW2}), and  we
need to perform one integration over $\kk_1$ or over $\kk_2$ to
dispose of the $\delta$-functions.
By calling simply $\qq$ the remaining integration
variable we obtain
\begin{align}
    \frac{dW_h^{(2)}}{d^2p_t}(x,b,\beta) = &\ \Gamma_h(x,b-\beta)
        \int \Gamma_A(x'_1,b) \Gamma_A(x'_2,b) dx'_1dx'_2
        \non  \\
        &\times \int d^2q \Big[
        \sigma(\qq) \sigma(\ppt-\qq)
        - 2 \sigma(\qq) \sigma(\ppt) \Big]
  \label{2sc}
\end{align}
As discussed in Section \ref{sec:cancdiv}, the negative term in
the expression above subtracts the leading inverse power divergence in
the integrand leaving only a logarithmic divergence. However, the
cancellation happens only after performing the integral over $\qq$,
which may be a difficult result to achieve numerically (actually this is
not a problem for the two-scattering term, due to the low
dimensionality of the integral, but becomes a big issue from three
scatterings on).

There are two divergences to be subtracted: one in $\qq \sim 0$ and the other
in $\qq \sim \ppt$, but the subtraction term is divergent just in
$\qq \sim 0$, and the cancellation of the inverse power singularities
is obtained only after performing the integration over $\qq$.
To allow the numerical integration to do a better and
faster job, we want that the divergences in the convolution
term and in the subtraction term be cancelled directly in the
integrand. This is obtained by symmetrizing the integrand with respect to an
interchange of the two singularities in the convolution term.
Let's introduce therefore an operator that performs the interchange of
the two singularities:
\begin{equation*}
    \mathbb{T} \ : \ \qq \ \ra \ \ppt - \qq \ ,
\end{equation*}
so that
\begin{equation*}
    \mathbb{T} \int\di^2q \, f(\qq) = \int\di^2q \, f(\ppt-\qq) \ .
\end{equation*}
Note that the change of variables operated by $\mathbb{T}$ has unit Jacobian
and that $\mathbb{T}^2=\mathbb{I}$. Then, we define the symmetrized
two-scattering term as
\begin{align*}
    \frac{dW_{A}^{(2)}}{d^2p_t}\In{sym} = \mathbb{S}^{(2)}
        \frac{dW_h^{(2)}}{d^2p_t} \ ,
\end{align*}
where we introduced the symmetrization operator
\begin{align}
     \mathbb{S}^{(2)} = \frac12 (\mathbb{I} + \mathbb{T}) \ .
  \label{s2}
\end{align}
The result is:
\begin{align}
    \frac{dW_{A}^{(2)}}{d^2p_t}\In{sym}(x,b,\beta) \ = & \ \Gamma_h(x,b-\beta)
        \int \Gamma_A(x'_1,b) \Gamma_A(x'_2,b) dx'_1dx'_2
        \non  \\
        & \times  \int d^2q \Big[
        \sigma(\qq) \sigma(\ppt-\qq)
        - \sigma(\qq) \sigma(\ppt)
        - \sigma(\ppt-\qq) \sigma(\ppt) \Big] \ .
  \label{2scsym}
\end{align}
Note that the first term in (\ref{2scsym}) describes two subsequent
scatterings with total transverse momentum $p_t$ and is the naive pQCD
result. The two negative terms are the absorption
terms induced by probability conservation.
The two IR divergences of the first term are canceled by these
two subtraction terms: as $\qq \ra \bf{0}$ by the first one and as
$\qq \ra \ppt$ by the second one. The remaining linear singularity
gives a zero contribution because it is odd in a neighborhood of
$\qq=0$ and $\qq=\ppt$ so that only the logarithmic divergence remain.
Note that now the two divergences are subtracted directly in the
integrand, which was the goal of the symmetrization procedure. \\
Eq.~(\ref{2scsym}) is the expression that we use in the numerical
computations of the transverse spectrum at high $p_t$.
It could have been guessed directly from Eq.~(\ref{2sc}), but the use of the
symmetrization operator (\ref{s2}) will facilitate the discussion of
the more complicated three scattering term.

\subsection{Three-scattering term}

To prepare the ground for the treatment of the three-scattering term,
we note that $\mathbb{T}$ generates the group of the
permutations of the two singularities $\qq\sim 0$ and $\qq\sim\ppt$; this is
called the symmetric group of order 2 and indicated as
$S_2=\langle\mathbb{T}\rangle=\{\mathbb{I},\mathbb{T}\}$, where
$\langle\mathbb{T}\rangle$ means ``generated by $\mathbb{T}$''. It's then easy
to see that we can construct the symmetrizing operator (\ref{s2}) by summing
all the elements of $S_2$ and by dividing by its cardinality.

From Eq.~(\ref{dW3}), after exploiting the $\delta$-functions, the
three-scattering term reads
\begin{align}
    \frac{dW_h^{(3)}}{d^2p_t} & (x,b,\beta) = \Gamma_h(x,b-\beta)
        \int \Gamma_A(x'_1,b) \Gamma_A(x'_2,b) \Gamma_A(x'_3,b)
        dx'_1dx'_2dx'_3  \\
    & \times \frac1{3!} \int \di^2 q \di^2 r
        \left[ \sigma(\qq) \sigma(\rr) \sigma(\ppt-\qq-\rr)
        - 3 \sigma(\qq) \sigma(\ppt-\qq)  \sigma(\ppt)
        + 3 \sigma(\qq) \sigma(\rr)  \sigma(\ppt) \right] .
  \label{3sc}
\end{align}
Following the general analysis previously done at the end of the last
paragraph, we observe that in (\ref{3sc}) in absence of the cutoff we
would have four divergences, i.e:
\begin{align}
    \qq \sim 0, \hspace*{.7cm} \rr \sim 0, \hspace*{.7cm}
    \ppt - \qq - \rr \sim 0, \hspace*{.7cm} \ppt-\qq \sim 0
  \label{divin3}
\end{align}
Then, to write the symmetrized three-scattering term, we need to consider the
group $S_4$ of the permutations of these four divergences, which has $4!=24$
elements:
\begin{align*}
    {\cal P}_{Bsym}^{(3)} = \mathbb{S}^{(3)} {\cal P}_{B}^{(2)}
\end{align*}
where
\begin{align*}
     \mathbb{S}^{(3)} = \frac{1}{24} \sum_{\ \mathbb{T}\in S_4} \mathbb{T}
\end{align*}
When applying this operator to the three-scattering term the resulting
expression has 49 terms and is too long to be discussed
here. To have an idea of the result, we will consider only the
subgroup $S_3$ given by the permutations of the first three
divergences in (\ref{divin3}), which are the divergences that appear in the
first term of (\ref{3sc}), i.e. the naive three-scattering term. After the
symmetrization it will be immediate to check that all the ``single''
divergences cancel explicitly in the integrand, while ``double''
divergences cancel only after performing the integrations over the
transverse momenta. We call ``single'' divergence a point $(\qq,\rr)$
such that only one of the expressions in (\ref{divin3}) is near zero,
and ``double'' divergence a point such that two of these terms are
nearly zero. For example $\{ \qq \sim 0 ;\ \rr \not\sim 0,\ppt,\ppt-\qq \}$
and $\{ \qq \sim 0 ;\ \rr \sim \ppt\}$
are respectively a single and a double divergence.

The first step is the definition of the operators that
exchange the three singularities:
\begin{align*}
    \mathbb{T}_1 \ : \ \left\{ \bay{l}
        \qq \ \ra \ \rr  \\
        \rr \ \ra \ \qq  \eay \right. \ \ \
    \mathbb{T}_2 \ : \ \left\{ \bay{l}
        \qq \ \ra \ \ppt - \qq - \rr  \\
        \rr \ \ra \ \rr  \eay \right. \ \ \
    \mathbb{T}_3 \ : \ \left\{ \bay{l}
        \qq \ \ra \ \qq  \\
        \rr \ \ra \ \ppt - \qq - \rr \eay \right. 
\end{align*}
Note that they are idempotent: $\mathbb{T}_i = \mathbb{I}$. 
Next, we observe that the group $S_3$ of the permutations of the three
singularities is made of $3!=6$ objects, and that
\begin{align*}
    S_3=\langle\mathbb{T}_1,\mathbb{T}_2,\mathbb{T}_3\rangle
    = \{\mathbb{T}_0,\mathbb{T}_1,\mathbb{T}_2,\mathbb{T}_3,
    \mathbb{T}_4,
    \mathbb{T}_5 \} \ ,
\end{align*}
where $\mathbb{T}_0=\mathbb{I},\ \mathbb{T}_4=\mathbb{T}_1\mathbb{T}_2$ and
$\mathbb{T}_5=\mathbb{T}_1\mathbb{T}_3 $, so that the reduced
symmetrizing operator is
\begin{equation*}
     \mathbb{S}^{(3)}_{red} = \frac1{3!} \sum_{i=0}^5 \mathbb{T}_i \ .
\end{equation*}
Finally one can write the partially symmetrized three-scattering probability:
\begin{align}
    \frac{dW_{A}^{(3)}}{d^2p_t}\In{sym} (x,b,\beta) &
        =  \mathbb{S}^{(3)}_{red}
        \frac{dW_{A}^{(3)}}{d^2p_t}(x,b,\beta) = \non \\
    =  \Gamma_h(x,b-\beta) & \
        \int \Gamma_A(x'_1,b) \Gamma_A(x'_2,b) \Gamma_A(x'_3,b)
        dx'_1dx'_2dx'_3 d^2k_1 d^2k_2 d^2k_3
        \non \\
    \times  \frac1{3!} & \ds \int \di x^\prime \di^2 q \di^2r
        \Big[ \sigma(\qq) \sigma(\rr) \sigma(\ppt-\qq-\rr)
        \non \\
    & - \frac12 \sigma(\qq) \sigma(\rr) \sigma(\ppt-\qq)
        + \frac12 \sigma(\qq) \sigma(\ppt-\qq) \sigma(\ppt)
        \non \\
    & - \frac12 \sigma(\qq) \sigma(\rr) \sigma(\ppt-\rr)
        + \frac12 \sigma(\ppt-\rr) \sigma(\qq) \sigma(\ppt)
        \non \\
    & - \frac12 \sigma(\ppt-\qq-\rr) \sigma(\qq) \sigma(\ppt-\qq)
        + \frac12 \sigma(\ppt-\rr) \sigma(\ppt-\qq) \sigma(\ppt)
        \non \\
    & - \frac12 \sigma(\ppt-\qq-\rr) \sigma(\rr) \sigma(\ppt-\rr)
        + \frac12 \sigma(\ppt-\rr) \sigma(\ppt-\qq-\rr) \sigma(\ppt)
        \non \\
    & - \frac12 \sigma(\rr) \sigma(\ppt-\qq-\rr) \sigma(\qq+\rr)
        + \frac12 \sigma(\rr) \sigma(\qq-\rr) \sigma(\ppt)
        \non \\
    & - \frac12 \sigma(\qq) \sigma(\ppt-\qq-\rr) \sigma(\qq+\rr)
        + \frac12 \sigma(\qq) \sigma(\qq-\rr) \sigma(\ppt)\Big] \ .
  \label{3scsym}
\end{align}
Analogously to what has been done for the two-scattering term, one can
see by inspection that the four single divergences \eqref{divin3}
explicitly cancel in the
integrand, while double divergences cancel only
after performing the integrations over $q$ and $r$. By considering all four
singularities, and by using the whole $S_4$ group we would get
explicit cancellation of both ``single'' and ``double'' divergences
directly in the integrand. Nonetheless, the partial symmetrization is
enough to get satisfactory numerical results.

In conclusion, to compute numerically the expansion of the transverse
minijet spectrum in the number of scatterings one has to fully exploit
the symmetry properties of each term, in such a way that all the
divergences get cancelled directly in the integrand. This is crucial
to obtain a good numerical precision and to speed up the computation
of the terms with three or more scatterings. In this appendix we
developed a general technique to perform such a symmetrization.

\end{appendix}


\end{document}